\documentclass[usenatbib]{mn2e}
\usepackage{color}
\usepackage{graphicx}
\usepackage{amssymb}
\usepackage{amsmath}
\usepackage{aas_macros}
\usepackage{natbib}
\usepackage{amsmath}
\usepackage{graphicx}
\usepackage{color}
\usepackage{mathrsfs}
\usepackage{epsfig}
\newcommand{\pasa}{PASA}

\title[LAEs with EW$_{\rm 0}$(Ly$\alpha$) $\simeq 200-400$\AA\ at $z\sim 2$]
 {
Ly$\alpha$ Emitters with Very Large Ly$\alpha$ Equivalent Widths, 
EW$_{\rm 0}$(Ly$\alpha$) $\simeq 200-400$\AA, at $z\sim 2$
}
\author[T. Hashimoto et al.]
{
Takuya~Hashimoto$^{1,2}$, 
Masami~Ouchi$^{3,4}$, 
Kazuhiro~Shimasaku$^{1,5}$,  
Daniel~Schaerer$^{6}$,
\newauthor
Kimihiko~Nakajima$^{6,7}$, 
Takatoshi~Shibuya$^{3}$, 
Yoshiaki~Ono$^{3}$, 
Michael~Rauch$^{8}$,
\newauthor 
and Ryosuke~Goto$^{1}$. 
\newauthor
\\
$^1$
Department of Astronomy, Graduate School of Science, The University of Tokyo, Tokyo 113-0033, Japan\\
$^2$
Universit\'e de Lyon, Lyon, F-69003; Universit\'e de Lyon 1, 
Observatoire de Lyon, 9 avenure Charles Andr\'e, Saint-Genis Laval, F-69230;\\ 
CNRS, UMR 5574, Centre de Recherche Astrophysique de Lyon; 
Eclode Normale Sup\'erieure de Lyon, Lyon, F-69007, France\\
$^3$
Institute for Cosmic Ray Research, The University of Tokyo, 5-1-5 Kashiwanoha, Kashiwa, Chiba 277-8582, Japan\\
$^4$
Kavli Institute for the Physics and Mathematics of the Universe (WPI),The University of Tokyo, 5-1-5 Kashiwanoha, \\Kashiwa, Chiba 277-8583, Japan\\
$^5$
Research Center for the Early Universe, Graduate School of Science, The University of Tokyo, Tokyo 113-0033, Japan\\
$^6$
Observatoire de Gen\`eve, Universit\'e de Gen\`eve, 51 Ch. des Maillettes, 1290 Versoix, Switzerland\\
$^7$
European Southern Observatory, Karl-Schwarzschild-Stra\ss e 2, 85748 Garching, Germany\\
$^8$
Observatories of the Carnegie Institution of Washington, 813 Santa Barbara Street, Pasadena, CA 91101, USA
}

\date{Accepted 2016 November 1. Received 2016 November 1; in original form 2016 September 6}

\pagerange{\pageref{firstpage}--\pageref{lastpage}} \pubyear{2016}

\begin{document}

\label{firstpage}

\maketitle

\begin{abstract}
We present physical properties of spectroscopically confirmed Ly$\alpha$ emitters (LAEs) 
with very large rest-frame Ly$\alpha$ equivalent widths EW$_{\rm 0}$(Ly$\alpha$). 
Although the definition of large EW$_{\rm 0}$(Ly$\alpha$) LAEs is usually difficult 
due to limited statistical and systematic uncertainties, 
we identify six LAEs selected from $\sim 3000$ LAEs at $z\sim 2$ 
with reliable measurements of EW$_{\rm 0}$ (Ly$\alpha$) $\simeq 200-400$ \AA\ 
given by careful continuum determinations with
our deep photometric and spectroscopic data.
These large EW$_{\rm 0}$(Ly$\alpha$) LAEs do not have signatures of AGN,
but notably small stellar masses of $M_{\rm *} = 10^{7-8}$ $M_{\rm \odot}$
and high specific star-formation rates 
(star formation rate per unit galaxy stellar mass)
of $\sim 100$ Gyr$^{-1}$.
These LAEs are characterized by the median values of
$L({\rm Ly\alpha})=3.7\times 10^{42}$ erg s$^{-1}$ and
$M_{\rm UV}=-18.0$ as well as
the blue UV continuum slope of $\beta = -2.5\pm0.2$
and
the low dust extinction $E(B-V)_{\rm *} = 0.02^{+0.04}_{-0.02}$,
which indicate a high median Ly$\alpha$ escape fraction of
$f_{\rm esc}^{\rm Ly\alpha}=0.68\pm0.30$. 
This large $f_{\rm esc}^{\rm Ly\alpha}$ value is explained by
the low {\sc Hi} column density in the ISM that is consistent with 
FWHM of the Ly$\alpha$ line, ${\rm FWHM (Ly\alpha)}=212\pm32$ km s$^{-1}$, 
significantly narrower than those of small EW$_{\rm 0}$(Ly$\alpha$) LAEs. 
Based on the stellar evolution models,
our observational constraints of
the large EW$_{\rm 0}$ (Ly$\alpha$), the small $\beta$, and
the rest-frame He{\sc ii} equivalent width
imply that at least a half of our large EW$_{\rm 0}$(Ly$\alpha$) LAEs
would have young stellar ages of $\lesssim 20$ Myr and very low metallicities
of $Z<0.02 Z_\odot$ 
regardless of the star-formation history.
\end{abstract}


\section{INTRODUCTION} \label{sec:introduction}

Photometric studies of Ly$\alpha$ emitters 
(LAEs: \citealt{cowie1998, rhoads2000, ouchi2003, malhotra2004, gronwall2007}) 
have revealed that 
about $4-10\%$ ($10-40\%$) of LAEs at $z\sim2-3$ ($z\sim4-6$) show 
extremely large rest-frame Ly$\alpha$ equivalent widths, 
EW$_{\rm 0}$ (Ly$\alpha$) $\gtrsim 200$ \AA\ 
($z\sim2-3$: \citealt{nilsson2007, mawatari2012}, 
$z\sim4-6$: \citealt{malhotra2002, shimasaku2006, ouchi2008, zhenya_zheng2014}). 
Several spectroscopic studies have also identified LAEs with large EW$_{\rm 0}$ (Ly$\alpha$) 
values (\citealt{dawson2004, wang2009, adams2011, kashikawa2012}). 

\cite{schaerer2003} and \cite{raiter2010} have constructed   
stellar evolution models
that cover various metallicities ($Z=0-1.0 Z_{\rm \odot}$) 
and a wide range of initial mass functions (IMFs). 
According to the models of \cite{schaerer2003} and \cite{raiter2010}, 
the value of EW$_{\rm 0}$ (Ly$\alpha$) $\gtrsim 200$ \AA\ can be explained by 
stellar populations with 
a very young stellar age ($\lesssim 10$ Myr), a very low metallicity, or a top-heavy IMF 
(cf., \citealt{charlot1993, malhotra2002}). 
Thus, large EW$_{\rm 0}$(Ly$\alpha$) LAEs are particularly interesting 
as candidates for galaxies at an early stage of the galaxy formation 
or galaxies with an exotic metallicity/IMF (\citealt{schaerer2002}). 
The models of \cite{schaerer2003} and \cite{raiter2010} have shown that 
the He{\sc ii} $\lambda 1640$ line 
is an useful indicator to break the degeneracy between the stellar age and metallicity. 
This is due to the fact that the high excitation level of He{\sc ii}, $h\nu = 54.4$ eV, 
can be achieved only by massive stars with extremely low metallicities 
($Z\sim 0 - 5 \times10^{-6}$ $Z_{\rm \odot}$). 
These models predict that 
galaxies hosting zero-metallicity stars (Population {\sc III} stars; hereafter Pop {\sc III}) 
can emit He{\sc ii} whose rest-frame EW, EW$_{\rm 0}$(He{\sc ii}), is up to a few times 10 \AA. 
The UV continuum slope ($\beta$), defined as $f_{\lambda} \propto \lambda^{\beta}$, 
is also powerful to place constraints on the stellar age and metallicity 
because the $\beta$ value ranges to as low as $\gtrsim -3.0$ 
depending on the stellar age and metallicity (\citealt{schaerer2003, raiter2010}). 
Therefore, it is important to simultaneously examine 
EW$_{\rm 0}$(Ly$\alpha$), EW$_{\rm 0}$(He{\sc II}), and $\beta$ values 
to put constraints on stellar ages and metallicities of 
large EW$_{\rm 0}$(Ly$\alpha$) LAEs. 

There are two problems in previous large EW$_{\rm 0}$(Ly$\alpha$) LAE studies. 
First, EW$_{\rm 0}$(Ly$\alpha$) measurements have large uncertainties. 
Because LAEs are generally faint in continua, 
it is difficult to measure the continuum flux at $1216$ \AA\ from spectroscopic data. 
Thus, most LAE studies have estimated the continuum flux at $1216$ \AA\ 
from photometric data in the wavelength range redward of 
$1216$ \AA. 
Furthermore, previous studies have assumed the flat UV continuum slope, $\beta=-2.0$, 
to estimate the continuum flux at $1216$ \AA. 
Since large EW$_{\rm 0}$(Ly$\alpha$) LAEs are typically very faint 
in the continuum (\citealt{ando2006}), 
large uncertainties remain in EW$_{\rm 0}$(Ly$\alpha$) values 
even if the continuum fluxes at $1216$ \AA\ are derived from photometric data. 
%
Second, detailed physical properties of large EW$_{\rm 0}$(Ly$\alpha$) LAEs 
have been scarcely investigated. 
There are no studies that placed constraints on stellar ages and metallicities of 
large EW$_{\rm 0}$(Ly$\alpha$) LAEs based on 
EW$_{\rm 0}$(Ly$\alpha$), EW$_{\rm 0}$(He{\sc ii}), and $\beta$ values. 
\cite{kashikawa2012} have examined the stellar age and metallicity 
of a large EW$_{\rm 0}$(Ly$\alpha$) LAE at $z\sim6.5$ 
based on EW$_{\rm 0}$(Ly$\alpha$) and EW$_{\rm 0}$(He{\sc ii}) values. 
However, the result is practically based on the EW$_{\rm 0}$(Ly$\alpha$) value 
because the EW$_{\rm 0}$(He{\sc ii}) value is only an upper limit.

In this study, we examine physical properties of six 
large EW$_{\rm 0}$(Ly$\alpha$) LAEs that are spectroscopically confirmed at $z\sim2$. 
By modeling deep FUV photometric data with no apriori assumption on $\beta$, 
we carefully estimate EW$_{\rm 0}$(Ly$\alpha$) and $\beta$ values of our LAEs. 
Remarkably, we find that our LAEs have large EW$_{\rm 0}$(Ly$\alpha$) values 
ranging from 
$160$ to $357$ \AA\ with a mean value of $252\pm30$ \AA. 
The $\beta$ values of our LAEs vary from 
$1.6$ to $-2.9$ 
with a small median value of $-2.5\pm0.2$. 
In order to place constraints on stellar ages and metallicities 
of our large EW$_{\rm 0}$(Ly$\alpha$) LAEs, 
we compare observational constraints of the large EW$_{\rm 0}$(Ly$\alpha$), 
the small $\beta$, and EW$_{\rm 0}$(He{\sc ii}) 
with theoretical models of \cite{schaerer2003} and \cite{raiter2010}. 
Since these theoretical models have fine metallicity grids at the low metallicity range, 
we can investigate stellar ages and metallicities 
of our large EW$_{\rm 0}$(Ly$\alpha$) LAEs in detail. 
We also derive physical quantities such as 
the stellar mass ($M_{\rm *}$), the star formation rate (SFR), 
the full width at the half maximum, FWHM, of the Ly$\alpha$ lines, 
and the Ly$\alpha$ escape fraction ($f^{\rm Ly\alpha}_{\rm esc}$) 
from the spectral data and photometric data (SED fitting).

This paper is organized as follows. 
We describe our large EW$_{\rm 0}$(Ly$\alpha$) LAE sample 
and data in \S \ref{sec:sample_data}.
In \S \ref{sec:results}, we derive EW$_{\rm 0}$(Ly$\alpha$) and $\beta$ values 
as well as several observational quantities of our LAEs.
A discussion in the context of physical properties of large EW$_{\rm 0}$(Ly$\alpha$) LAEs 
is given in \S \ref{sec:discussion}, 
followed by conclusions in \S \ref{sec:conclusions}. 
Throughout this paper, magnitudes are given in the AB system
\citep{oke1983}, and we assume a $\Lambda$CDM cosmology 
with $\Omega_{\small m} = 0.3$, $\Omega_{\small \Lambda} = 0.7$
and $H_{\small 0} = 70$ km s$^{-1}$ Mpc$^{-1}$.

\section{Sample and Data} \label{sec:sample_data}

\subsection{Large EW$_{\rm 0}$(Ly$\alpha$) LAE Sample} \label{subsec:sample}

Our large EW$_{\rm 0}$(Ly$\alpha$) LAEs are taken 
from the largest ($N\sim3000$) parent LAE sample 
at $z\sim2.2$ spanning in the COSMOS field, the Chandra Deep Field South (CDFS), 
and the Subaru/{\it XMM-Newton} Deep Survey (SXDS) 
(\citealt{nakajima2012, nakajima2013, konno2016}; Kusakabe et al. in prep.). 
The parent sample is based on 
Subaru/Suprime-Cam imaging observations 
with our custom made narrow band filter, NB387. 
The central wavelength and the FWHM of NB387 are $3870$ \AA\ and $94$ \AA, respectively. 
The parent LAE sample has been selected by the following color criteria:  
\begin{eqnarray}
u^{*} - NB387 > 0.5 \ \& \ B - NB387 > 0.2, 
\end{eqnarray}
satisfying the condition that the EW$_{\rm 0}$(Ly$\alpha$) value 
should be larger than $30$ \AA. 
The parent sample has been used to examine LAEs' 
metal abundances and ionization parameters (\citealt{nakajima2012, nakajima2013, nakajima2014}), 
kinematics of the inter-stellar medium (ISM) (\citealt{hashimoto2013, shibuya2014b, hashimoto2015}), 
diffuse Ly$\alpha$ haloes (\citealt{momose2014, momose2016}), 
morphologies (\citealt{shibuya2014a}), 
dust properties (\citealt{kusakabe2015}), 
and the Ly$\alpha$ luminosity function (\citealt{konno2016}).

From the parent sample, we use six objects with strong NB387 excesses, 
\begin{eqnarray}
u^{*} - NB387 > 1.0 \ \& \ B - NB387 > 1.4, 
\end{eqnarray}
as well as Ly$\alpha$ identifications that are listed in Table \ref{tab:sample}: 
four from the COSMOS field, 
COSMOS-08501, COSMOS-40792, COSMOS-41547, and COSMOS-44993 
and 
two from the SXDS-center (SXDS-C) field, 
SXDS-C-10535 and SXDS-C-16564. 
Since our targets are large EW$_{\rm 0}$(Ly$\alpha$) objects 
whose Ly$\alpha$ emission originates from star-forming activities, 
we examine if our sample includes a Ly$\alpha$ Blob 
(LAB: \citealt{moller1998}, \citealt{steidel2000}). 
This is because Ly$\alpha$ emission of LABs is thought to be powered by 
AGN activities (e.g., \citealt{haiman2001}), superwinds from starburst galaxies (e.g., \citealt{taniguchi2000}), and cold accretion (e.g., \citealt{haiman2000}). 
To check the presence of LABs, we have inspected the isophotal areas of NB387 images 
that trace the Ly$\alpha$ morphologies. 
We have obtained 
$4.2$ arcsec$^{2}$ (COSMOS-08501), 
$1.1$ arcsec$^{2}$ (COSMOS-40792),
$1.7$ arcsec$^{2}$ (COSMOS-41547), 
$1.5$ arcsec$^{2}$ (COSMOS-44993), 
$1.7$ arcsec$^{2}$ (SXDS-C-10535), 
and 
$6.2$ arcsec$^{2}$ (SXDS-C-16564). 
These isophotal areas correspond to the radii of $9-21$ kpc at $z\sim2.2$. 
These radii are spatially compact compared to the half light radii of typical $z\sim3$ LABs, 
$30-300$ kpc (\citealt{steidel2000, matsuda2004}). 
Thus, we conclude that our large EW$_{\rm 0}$(Ly$\alpha$) LAEs 
do not include a  LAB.

\begin{table*}
\centering
\caption{Sample of Large EW$_{\rm 0}$(Ly$\alpha$) LAEs}
\label{tab:sample}
\begin{tabular}{cccccccc}
\hline
{Object} & {$\alpha$(J2000)} & {$\delta$(J2000)} 
& {$u^{*} - NB387$} & {$B - NB387$} & {Line} & {$z_{\rm Ly\alpha}$} & {Source$^{a}$}\\
{(1)} & {(2)} & {(3)}  & {(4)} & {(5)} & {(6)} & {(7)} & {(8)}\\
\hline
COSMOS-08501 & 10:01:16.80 & +02:05:36.3 &  $1.45$ & $2.17$ & Ly$\alpha$ (MagE), H$\alpha$ (NIRSPEC) & $2.162$ & N13, H15\\
\hline
COSMOS-40792 & 09:59:46.66 & +02:24:34.2 &  $1.44$ & $2.02$ & Ly$\alpha$ (LRIS) & $2.209$ & S14\\
\hline
COSMOS-41547 & 09:59:41.91 & +02:25:00.0 &  $1.00$ & $1.90$ & Ly$\alpha$ (LRIS) & $2.152$ & S14\\
\hline
COSMOS-44993 & 09:59:53.87 & +02:27:11.0 &  $1.42$ & $1.48$ & Ly$\alpha$ (LRIS) & $2.214$ & S14\\
\hline
SXDS-C-10535 & 02:17:41.92 & -05:02:55.9 &  $1.11$ & $1.42$ & Ly$\alpha$ (LRIS) & $2.213$ & S14\\
\hline
SXDS-C-16564 & 02:19:09:54 & -04:57:13.3 &  $1.46$ & $1.95$ & Ly$\alpha$ (IMACS) & $2.176$ & N12\\
\hline
\end{tabular}
%
\begin{minipage}{170mm}
\begin{flushleft}
(1) Object ID;
(2) and (3) Right Ascension and Declination;
(4) and (5) $u^{*} - NB387$ and $B - NB387$ colors; 
(6) Spectroscopically identified line(s) and instruments used for observations;
(7) Redshifts inferred from the Ly$\alpha$ lines; 
and 
(8) Source of the information. 
\\
$^a$
N12: \cite{nakajima2012}; 
N13: \cite{nakajima2013}; 
S14: \cite{shibuya2014b}; 
H15: \cite{hashimoto2015}. 
\end{flushleft}
\end{minipage}
\end{table*}

\subsection{Photometric Data} \label{subsec:photo}

We performed photometry using SExtractor (\citealt{bertin1996}). 
We use $14$ bandpasses: $u^{*}, NB387, B, V, r', i'$, and $z'$ data 
taken with Subaru/Suprime-Cam, 
$J$ data taken with UKIRT/WFCAM, 
$H$ and $K$ data taken with CFHT/WIRCAM (UKIRT/WFCAM) for COSMOS (SXDS-C), 
and {\it Spitzer}/IRAC $3.6, 4.5, 5.8,$ and $8.0$ $\mu$m  
from the {\it Spitzer} legacy survey of the UDS fields. 

For the detailed procedure of photometry, we refer the reader to \cite{nakajima2012}. 
Recently, \cite{skelton2014} have re-calibrated zero-point magnitudes for 
the COSMOS and SXDS fields 
using 3D-HST (\citealt{brammer2012}) and 
CANDELS (\citealt{grogin2011,koekemoer2011}) data. 
\cite{skelton2014} have found that the zero-point magnitude offsets are from $0.00$ to $-0.25$. 
For secure estimates of physical quantities, 
we correct our zero-point magnitudes for the offsets 
listed in Tables $11$ and $12$ of  \cite{skelton2014}. 
Table \ref{tab:photometry} summarizes the photometry of our objects. 

\begin{table*}
\centering
\caption{Photometry of our LAEs}
\label{tab:photometry}
\begin{tabular}{lllllllllllllll}
\hline
{Object} & {$u^*$}& {NB387} & {\it{B}} & {\it{V}} & {\it{r'}} & {\it{i'}}& {\it{z'}} & {\it{J}} & {\it{H}} & {\it{K$_{s}$}} & {[3.6]}
& {[4.5]} & {[5.8]} & {[8.0]} \\
\hline
COSMOS &  &  &  &  &  &  &  &  &  &  &  &  &  & \\
\hline
08501 & 25.14 & 23.69 & 25.86 & 25.91 & 26.05 & 25.96 & 25.77 & 99.99 & 26.47 & 25.85 & 99.99 & 99.99 & 99.99 & 99.99\\ 
& (0.03) & (0.03) & (0.05) & (0.17) & (0.16) & (0.20) & (0.52) & (-) & (1.39) & (1.00) & (-) & (-) & (-) & (-) \\
\hline
40792 & 26.72 & 25.28 & 27.30 & 27.14 & 27.94 & 31.21 & 27.41 & 99.99 & 99.99 & 99.99 & 25.54 & 25.79 & 99.99 & 99.99\\
& (0.14) & (0.12) & (0.21) & (0.56) & (1.38) & (3.38) & (0.62) & (-) & (-) & (-) & (1.45) & (1.00) & (-) & (-)\\ 
\hline 
41547 & 26.06 & 25.07 & 26.97 & 26.59 & 26.70 & 26.21 & 27.46 & 99.99 & 25.71 & 24.94 & 99.99 & 99.99 & 22.50 & 22.72\\
& (0.08) & (0.10) & (0.15) & (0.32) & (0.30) & (0.25) & (0.68) & (-) & (0.66) & (1.42) & (-) & (-) & (0.62) & (2.18) \\ 
\hline 
44993 & 26.50 & 25.08 & 26.56 & 27.02 & 26.71 & 26.55 & 27.41 & 24.54 & 99.99 & 25.52 & 99.99 & 99.99 & 99.99 & 99.99\\
& (0.12) & (0.10) & (0.10) & (0.49) & (0.31) & (0.35) & (0.62) & (1.35) & (-) & (0.60) &  (-) & (-) & (-) & (-) \\
\hline
offset$^{a}$ & -0.16 & 0.00 & 0.03 & 0.23 & 0.20 & 0.12 & 0.21 & 0.08 & 0.07 & 0.07 & -0.02 & 0.03 & 0.05 & 0.12 \\ 
\hline
SXDS-C &  &  &  &  &  &  &  &  &  &  &  &  &  & \\ 
\hline
10535 & 25.84 & 24.73 & 26.15 & 26.29 & 26.64 & 26.61 & 26.94 & 27.12 & 27.54 & 25.64 & 25.92 & 99.99 & 99.99 & 99.99\\ 
& (0.08) & (0.10) & (0.08) & (0.12) & (0.22) & (0.22) & (0.83) & (0.62) & (1.78) & (0.72) & (1.45) & (-) & (-) & (-)\\
\hline
16564 & 24.14 & 22.68 & 24.63 & 24.83 & 24.97 & 25.09 & 25.16 & 25.18 & 23.83 & 24.40 & 25.49 & 29.91 & 99.99 & 99.99\\ 
& (0.02) & (0.01) & (0.02) & (0.03) & (0.05) & (0.05) & (0.14) & (0.37) & (0.19) & (0.20) & (1.11) & (4.24) & (-) & (-) \\
\hline 
offset$^{a}$ & -0.25 & 0.00 & 0.03 & 0.06 & 0.18 & 0.25 & 0.18 & -0.01 & -0.06 & -0.06 & -0.05 & 0.00 & -0.15 & -0.15 \\
\hline
\end{tabular}
%
\begin{minipage}{170mm}
\begin{flushleft}
All magnitudes are total magnitudes.
$99.99$ mag indicates a negative flux density.
Magnitudes in parentheses are $1\sigma$ uncertainties.

$^a$
Zero-point magnitude offsets quoted from \cite{skelton2014}. 
\end{flushleft}
\end{minipage}
\end{table*}

\subsection{Spectroscopic Data} \label{subsec:spec}

We carried out optical observations 
with Magellan/IMACS (PI: M. Ouchi), Magellan/MagE (PI: M. Rauch), 
and Keck/LRIS (PI: M. Ouchi). 
Details of the observations and data reduction procedures 
have been presented in 
\cite{nakajima2012} (IMACS), 
\cite{shibuya2014b} (LRIS), 
and \cite{hashimoto2015} (MagE). 
The spectral resolutions for our observations were 
$R\sim 700$ (IMACS), $\sim1100$ (LRIS), and $\sim4100$ (MagE). 
SXDS-C-16564 was observed with IMACS, 
from which we identified the Ly$\alpha$ line (\citealt{nakajima2012}).  
COSMOS-40792, COSMOS-41547, COSMOS-44933, and SXDS-C-10535 
were observed with LRIS. 
Although these LAEs are as faint as $B\sim26-27$, 
we detected the Ly$\alpha$ lines due to the high sensitivity of LRIS 
(\citealt{shibuya2014b}). 
COSMOS-08501 was observed with MagE, 
from which we identified the Ly$\alpha$ line (\citealt{hashimoto2015}). 
The H$\alpha$ line was also detected in COSMOS-08501 
with Keck/NIRSPEC at the significance level of $\sim5 \sigma$ (\citealt{nakajima2013}). 

\begin{figure*}
\includegraphics[width=13cm]{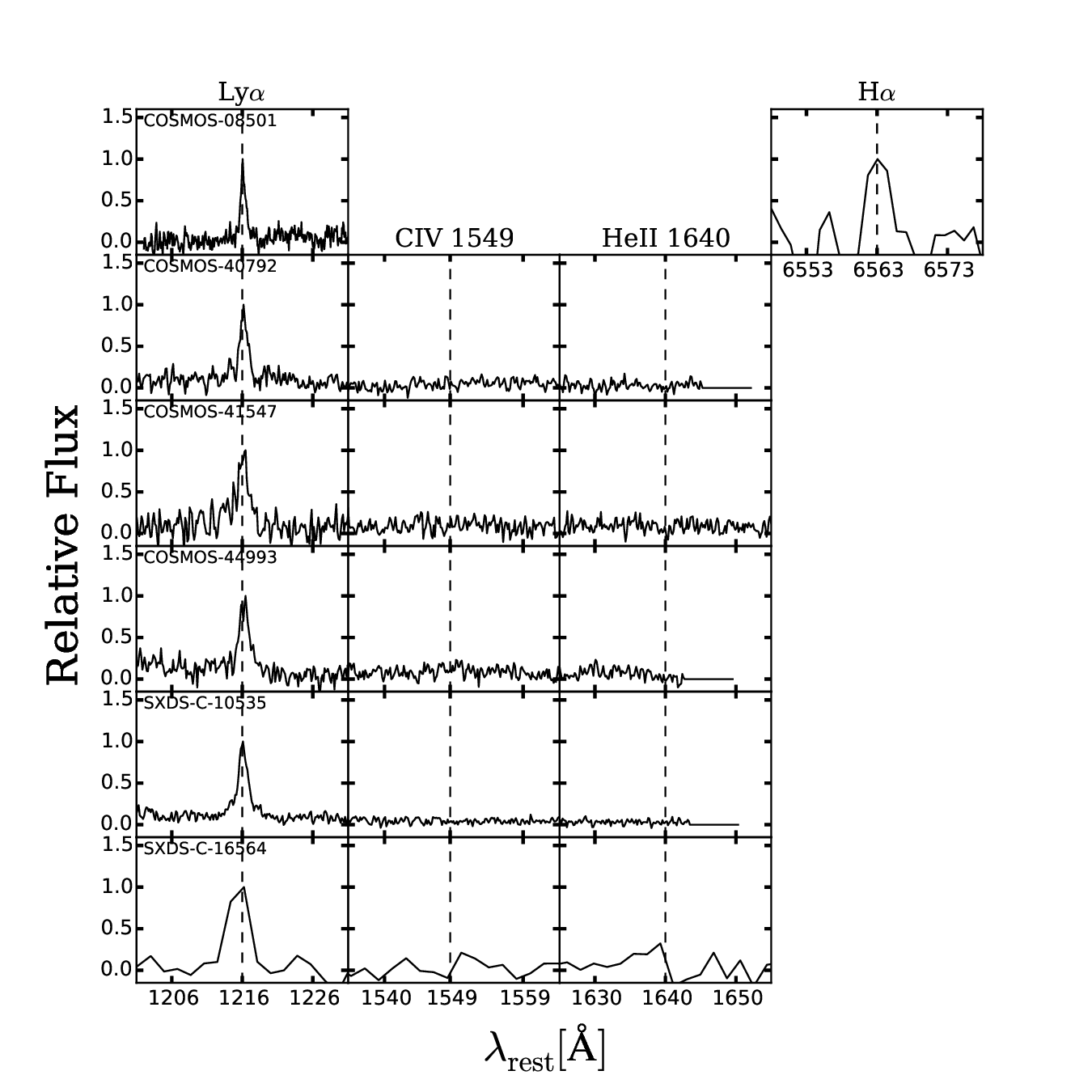}
\caption[]
{
From left to right, reduced 1D spectra corresponding to 
wavelength regions near 
Ly$\alpha$, C{\sc iv} $\lambda1549$, He{\sc ii} $\lambda1640$, 
and H$\alpha$ of our LAEs. 
The dashed lines in the 1D spectra show the expected locations of the lines. 
}
\label{fig:spectra}
\end{figure*}

We additionally search for the C{\sc iv} $\lambda1549$ and He{\sc ii} $\lambda1640$ lines 
in our LAEs. 
We determine a line to be detected, if there exists an emission line 
above the $3\sigma$ sky noise around the wavelength 
expected from the Ly$\alpha$ redshift. 
In this analysis, we measure the sky noise 
from the spectrum within $50$ \AA\ from the line wavelength. 
Neither C{\sc iv} nor He{\sc ii} was detected above $3\sigma$ in our LAEs. 
The flux upper limits of C{\sc iv} are used to 
diagnose signatures of AGN in our LAEs (\S \ref{subsec:agn}), 
while those of He{\sc ii} enable us to place constraints 
on the stellar ages and metallicities of our LAEs (\S \ref{subsec:age_metal}). 
Figure \ref{fig:spectra} shows 1D spectra corresponding to data 
around Ly$\alpha$, C{\sc iv}, He{\sc ii}, and H$\alpha$ lines. 

\subsection{AGN Activities in the Sample} \label{subsec:agn}

We examine whether our LAEs host an AGN in three ways. 
First, we compare the sky coordinates of the objects 
with those in very deep archival X-ray and radio catalogs (\citealt{elvis2009}). 
The sensitivity limits are 
$1.9\times19^{-16}$ ($0.5-2.0$ keV band), 
$7.3\times19^{-16}$ ($2-10$ keV band), 
and 
$5.7\times19^{-16}$ erg $^{-2}$ s$^{-1}$ ($0.5-10$ keV band). 
We also refer to the radio catalog constructed by \cite{schinnerer2010}. 
No counterpart for the LAEs is found in any of the catalogs. 

\begin{figure*}
\centering
\includegraphics[width=14cm]{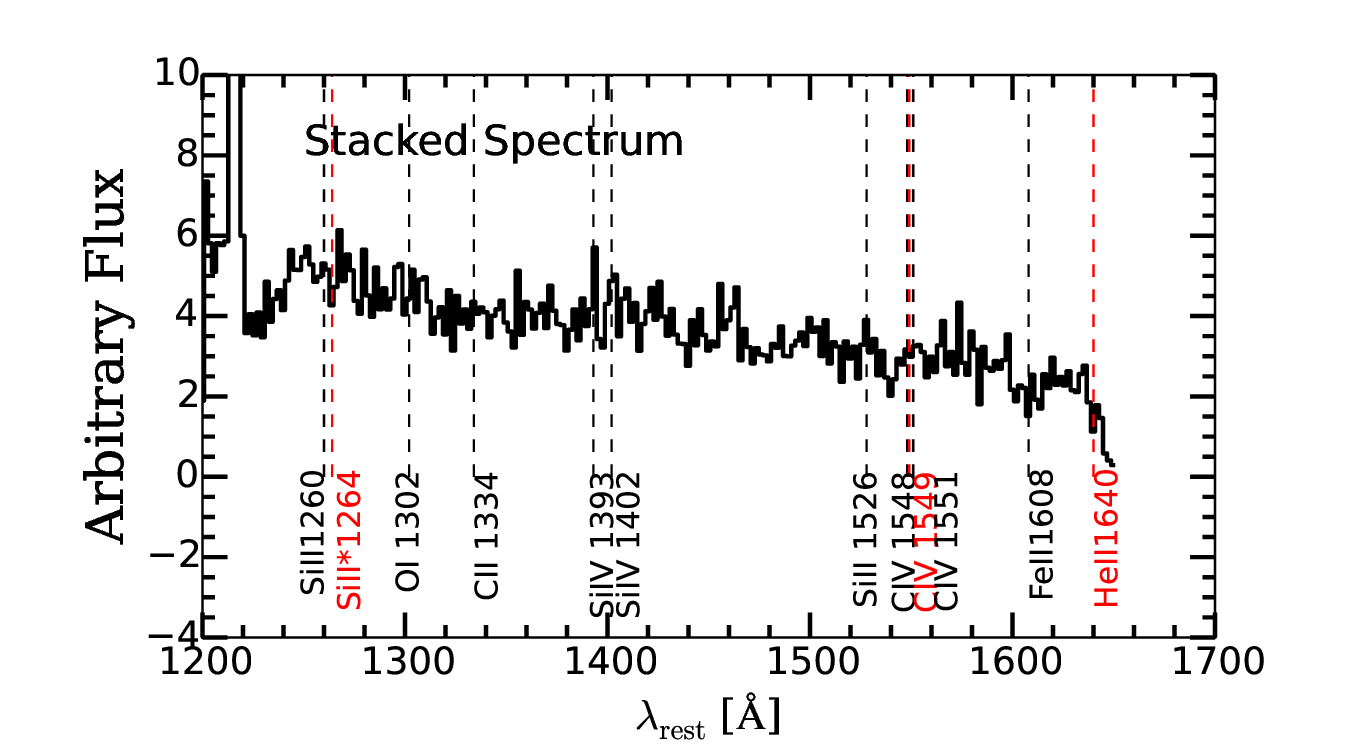}
\caption[]
{
Composite rest-frame UV spectrum of the four LRIS spectra. 
The black and red vertical dashed lines indicate wavelengths of 
interstellar absorption lines and emission lines, respectively. 
}
\label{fig:stacked_spectra}
\end{figure*}

Second, we search for the {\sc Civ} $1549$ line whose high ionization potential 
can be achieved by AGN activities. 
The {\sc Civ} line is not detected on an individual basis (\S \ref{subsec:spec}). 
To obtain a strong constraint on the presence of an AGN, 
we stack the four LRIS spectra 
by shifting individual spectral data from the observed to the rest frame. 
We infer the systemic redshifts of the four LRIS objects as follows. 
The Ly$\alpha$ line is known to be redshifted 
with respect to the systemic redshift by $200-400$ km s$^{-1}$ 
(e.g., \citealt{steidel2010, hashimoto2013, shibuya2014b, erb2014, henry2015, stark2015}). 
Based on an anti-correlation between the Ly$\alpha$ velocity offset 
and EW$_{\rm 0}$(Ly$\alpha$) (\citealt{hashimoto2013, shibuya2014b, erb2014}), 
we assume that our large EW$_{\rm 0}$(Ly$\alpha$) LAEs have 
the same Ly$\alpha$ velocity offsets as COSMOS-08501, 
$82\pm40$ km s$^{-1}$ (\citealt{hashimoto2015}). 
Figure \ref{fig:stacked_spectra} shows the stacked FUV spectrum of the four LRIS spectra. 
The {\sc Civ} line is not detected even in the composite spectrum. 
We obtain the $3\sigma$ lower limit of the flux ratio, 
$f_{\rm Ly\alpha}$/$f_{\rm CIV}$ $> 19.0$, 
where $f_{\rm Ly\alpha}$ and $f_{\rm CIV}$ are the Ly$\alpha$ and {\sc Civ} fluxes, 
respectively. 
The flux ratio is significantly larger than that for $z \sim 2-3$ radio galaxies, 
$f_{\rm Ly\alpha}$/$f_{\rm CIV}$ $= 6.9$  (\citealt{villar-martin2007}).

Finally, \cite{nakajima2013} have examined the position of COSMOS-08501 
in the BPT diagram (\citealt{baldwin1981}).
As shown in Figure $3$ of \cite{nakajima2013}, 
the upper limit of the flux ratio of [{\sc Nii}] $\lambda 6584$ and H$\alpha$, 
log([{\sc Nii}] $\lambda 6584$/H$\alpha$) $\lesssim -0.7$, 
indicates that COSMOS-08501 does not host an AGN. 

Thus, we conclude that no AGN activity is seen in our LAEs. 

\section{Results} \label{sec:results}

\subsection{SED Fitting} \label{subsec:sed_fit}

\begin{figure*}
\centering
\includegraphics[width=12cm]{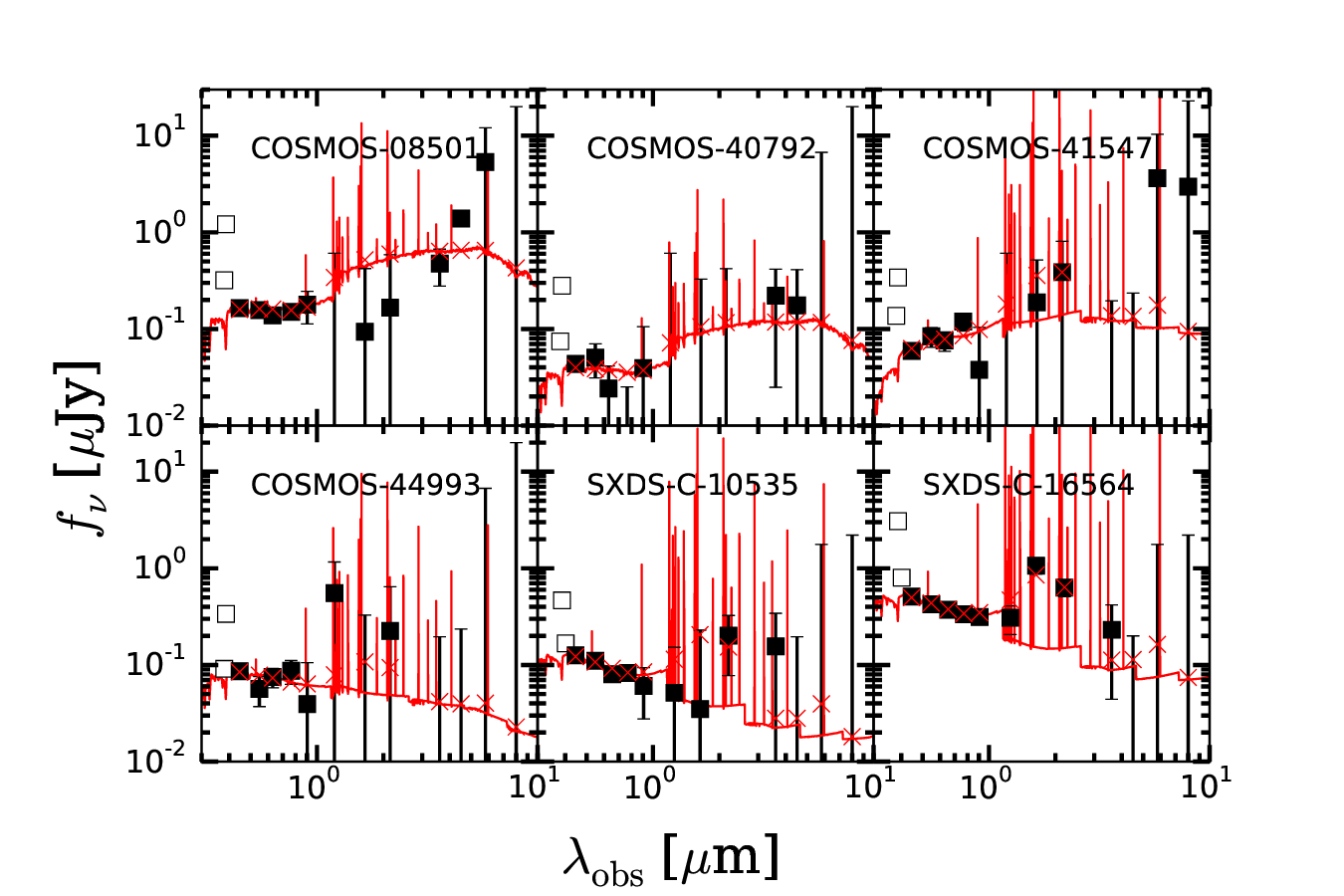}
\caption[]
{
Results of SED fitting for our LAEs. 
The filled squares denote the photometry points used for SED fitting, 
while the open squares are those omitted in SED fitting 
due to the contamination of Ly$\alpha$ emission and IGM absorption.
The red lines present the best-fit model spectra, 
while the red crosses correspond to 
the flux densities at individual passbands 
expected from the best-fit models. 
}
\label{fig:sed_fit}
\end{figure*}

We perform stellar population synthesis model fitting to our LAEs 
to derive the stellar mass ($M_{\rm *}$), stellar dust extinction ($E(B-V)_{\rm *}$), 
the stellar age, and the star-formation rate (SFR). 
For the detailed procedure, we refer the reader to \cite{ono2010b,ono2010a}. 
Briefly,  we use the stellar population synthesis model of GALAXEV
\citep{bc03} including nebular emission \citep{schaerer_de_barros2009}, 
and adopt the Salpeter IMF \citep{salpeter1955}. 
For simplicity, we use constant star formation models.
Indeed, several authors have assumed the constant star-formation history (SFH) 
for LAE studies at $z\sim2$ (e.g., \citealt{kusakabe2015, hagen2016}) 
and at $z>3$ (see Table 6 in \citealt{ono2010a}). 
Because LAEs are metal poor star-forming galaxies 
(\citealt{finkelstein2011, nakajima2012, nakajima2013, song2014}),
we choose a metallicity of $Z = 0.2$ $Z_{\odot}$. 
We use Calzetti's law \citep{calzetti2000} for $E(B-V)_{\rm *}$, 
and apply $18 \%$ IGM attenuation of 
continuum photons shortward of Ly$\alpha$ 
using the prescription of \cite{madau1995}. 
To derive the best-fit parameters, 
we use all bandpasses mentioned in \S \ref{subsec:photo} 
except for ${\it u^{*}}$ and NB387-band data. 
Neither ${\it u^{*}}$ nor NB387-band data have been used 
since the photometry of these bands is contaminated
by IGM absorption and/or Ly$\alpha$ emission.
Figure \ref{fig:sed_fit} shows the best-fit model spectra
with the observed flux densities. 
The derived quantities and their $1\sigma$ uncertainties 
are summarized in Table \ref{tab:sed_fit}. 

In Table \ref{tab:sed_fit}, our LAEs have stellar masses 
mostly $M_{\rm *}=10^{7-8} M_{\rm \odot}$ 
with a median value of $7.1^{+4.8}_{-2.8} \times 10^{7}M_{\rm \odot}$. 
The median value is smaller than that of $z\sim2$ LAEs with small EW$_{\rm 0}$(Ly$\alpha$), 
$2-5 \times 10^{8}M_{\rm \odot}$ (\citealt{nakajima2012, hagen2016}). 
\cite{nilsson2011, oteo2015} and \cite{shimakawa2016} have 
also studied stellar masses of $z\sim2$ LAEs. 
In these studies, there are no LAEs with $M_{\rm *} < 10^{8} M_{\rm \odot}$. 
These results indicate that our sample is consisted of low-mass LAEs.

The dust extinction of our LAEs varies from $E(B-V)_{\rm *} = 0.00$ to $0.25$ 
with a median value of $0.02^{+0.04}_{-0.02}$. 
This is lower than the typical dust extinction of $z\sim2$ LAEs, $E(B-V)_{\rm *} \sim 0.2 - 0.3$ 
(\citealt{guaita2011, nakajima2012, oteo2015}). 
This result shows that our LAEs have small amounts of dust.

\begin{table*}
\centering
\caption{Results of SED fitting}
\label{tab:sed_fit}
\begin{tabular}{cccccc}
\hline
{Object} & {$\chi^{2}$} & {log($M_{*}$)} & {$E(B-V)_{\rm *}$} 
& {log(age)} &{log(SFR)} \\
{} & {} & {($M_{\odot}$)}  & {} & {(yr)} & {($M_{\odot}$/yr)}\\
{(1)} & {(2)} & {(3)}  & {(4)} & {(5)} & {(6)}\\
\hline
COSMOS-8501 & 1.8 & $7.8^{+1.2}_{-0.3}$ & $0.08^{+0.04}_{-0.08}$ & $6.2^{+2.8}_{-1.1}$ & $1.68^{+1.24}_{-1.44}$\\
\hline
COSMOS-40792 & 3.8 & $8.9^{+0.2}_{-2.1}$ & $0.00^{+0.10}_{-0.00}$ & $9.4^{+0.0}_{-4.4}$ & $-0.37^{+2.53}_{-0.00}$\\
\hline
COSMOS-41547 & 4.8 & $8.1^{+0.4}_{-0.3}$ & $0.25^{+0.04}_{-0.07}$ & $6.9^{+0.8}_{-0.4}$ & $1.27^{+0.42}_{-0.51}$\\
\hline 
COSMOS-44993 & 3.1 & $7.6^{+1.2}_{-0.5}$ & $0.03^{+0.09}_{-0.03}$ & $7.4^{+1.6}_{-2.3}$ & $0.23^{+2.28}_{-0.28}$\\
\hline
SXDS1-10535 & 3.4 & $7.3^{+0.5}_{-0.0}$ & $0.00^{+0.02}_{-0.00}$ & $6.5^{+1.2}_{-1.4}$ & $0.80^{+1.50}_{-0.59}$\\
\hline
SXDS1-16564 & 9.2 & $7.9^{+0.0}_{-0.0}$ & $0.00^{+0.00}_{-0.00}$ & $6.5^{+0.1}_{-0.1}$ & $1.43^{+0.1}_{-0.1}$\\
\hline
\end{tabular}
%
\begin{minipage}{170mm}
\begin{flushleft}
Stellar metallicity is fixed to 0.2 $Z_{\odot}$.\\
(1) Object ID;
(2) $\chi^{2}$ of the fitting; 
(3) Stellar mass; 
(4) Stellar dust extinction;
(5) Stellar age; 
and 
(6) Star-formation rate.  
\end{flushleft}
\end{minipage}
\end{table*}

\subsection{Careful Estimates of EW$_{\rm 0}$(Ly$\alpha$) and $\beta$} \label{subsec:estimate_ew_beta}

We model a realistic FUV spectrum of a LAE to derive 
EW$_{\rm 0}$(Ly$\alpha$), the Ly$\alpha$ luminosity ($L({\rm Ly\alpha})$), 
the UV absolute magnitude ($M_{\rm UV}$), and $\beta$. 
As mentioned in \S \ref{sec:introduction}, 
EW$_{\rm 0}$(Ly$\alpha$) estimates in previous studies 
are based on several assumptions: 
(i) the UV continuum slope is flat, $\beta = -2.0$, 
and (ii) the pass-bands are ideal top-hat response functions
(e.g., \citealt{malhotra2002, guaita2011, mawatari2012}). 
These factors add systematic uncertainties in EW$_{\rm 0}$(Ly$\alpha$). 

In this work, the LAE spectrum is modeled as a combination of 
a delta-function Ly$\alpha$ line and a linear continuum, 
\begin{eqnarray}
f_{\rm \nu, Ly\alpha} 
= F({\rm Ly\alpha}) \times \delta(\nu - \nu_{\rm Ly\alpha}), 
\end{eqnarray}
\begin{eqnarray}
f_{\rm \nu, cont} 
= A \times \nu^{-(\beta_{\rm 1200-2800} + 2)},
\end{eqnarray}
where 
$f_{\rm \nu, Ly\alpha}$ ($f_{\rm \nu, cont}$ ) is the Ly$\alpha$ (continuum) 
flux per unit frequency in erg cm$^{-2}$ s$^{-1}$ Hz$^{-1}$, 
while $F({\rm Ly\alpha})$ is the integrated flux of the line in erg cm$^{-2}$ s$^{-1}$. 
The function $\delta(\nu - \nu_{\rm Ly\alpha})$ is a delta-function, 
and the function $A$ corresponds to the amplitude of the continuum flux. 
In equation (4), $A$ is expressed as 
\begin{eqnarray}
A = (2.0 \times 10^{15})^{\beta_{\rm 1200-2800}+2} \times 10^{-0.4(m_{\rm 1500} + 48.6)}, 
\end{eqnarray}
where $\beta_{\rm 1200-2800}$ is the UV continuum slope 
in the rest-frame wavelength range of $1200-2800$ \AA, 
while $m_{1500}$ indicates the apparent magnitude at 1500 \AA. 
With $f_{\rm \nu, cont}$ and $F({\rm Ly\alpha})$ in equations (3) and (4), 
the modeled flux in the $i$th band is defined as 
\begin{eqnarray}
f_{\rm \nu, model}^{(i)} 
= \frac{\int f_{\rm \nu} T_{\rm \nu}^{(i)} d\nu}{\int T_{\rm \nu}^{(i)} d\nu}= 
\end{eqnarray}
\begin{eqnarray}
\frac{
\int_{\nu_{\rm s}}^{\nu_{\rm Ly\alpha}} f_{\rm \nu, cont} T_{\rm \nu}^{(i)} d\nu
+ F({\rm Ly\alpha}) T_{\rm \nu, Ly\alpha}^{(i)}
+ \alpha \int_{\nu_{\rm Ly\alpha}}^{\nu_{\rm e}} f_{\rm \nu, cont} T_{\rm \nu}^{(i)} d\nu
 }
{\int_{\nu_{\rm s}}^{\nu_{\rm e}} T_{\rm \nu}^{(i)} d\nu}. 
\end{eqnarray}
In equation (7), 
$T_{\nu}^{(i)}$ is the response curve of the $i$th band. 
The constants of $\nu_{\rm s}$ and $\nu_{\rm e}$ indicate the frequencies 
corresponding to the upper and lower ends of the response curves, respectively. 
The constant $T_{\nu, Ly\alpha}^{(i)}$ is the response curve value 
of the $i$th band at the Ly$\alpha$ frequency, $\nu_{\rm Ly\alpha}$, 
where  $\nu_{\rm Ly\alpha}$ is 
calculated from the Ly$\alpha$ redshift, $z_{\rm Ly\alpha}$ (Table \ref{tab:sample}).
Finally, $\alpha$ means the continuum photon transmission 
shortward of Ly$\alpha$ after the IGM absorption. 
Using the prescription of \cite{madau1995}, at $z\simeq2$, it is 
\[
  \alpha = \begin{cases}
    0.82 & (\nu \geq \nu_{\rm Ly\alpha}) \\
    1.0 & (\nu < \nu_{\rm Ly\alpha}).
  \end{cases}
\]

To estimate EW$_{\rm 0}$(Ly$\alpha$) and other quantities, 
we compare the modeled flux in the $i$th band 
with the observed one in the $i$th band. 
The observed flux in the $i$th band is expressed as 
\begin{eqnarray}
f_{\rm \nu, obs}^{(i)} 
= 10^{-0.4({\rm AB^{(i)}} + 48.6)}, 
\end{eqnarray}
where ${\rm AB^{(i)}}$ is the AB magnitude of the $i$th band 
listed in Table \ref{tab:photometry}.
For each LAE, we use six rest-frame FUV data, from ${\it u^{*}}$ to $i'$-band. 
With equations (7) and (8), we search for the best-fitting spectrum that minimizes 
\begin{eqnarray}
\chi^{2} 
= \sum_{i}^{6}
\frac{\bigl \{ f^{(i)}_{\rm \nu, obs} - f^{(i)}_{\rm \nu, model} \bigl [ F(Ly\alpha), \beta_{\rm 1200-2800}, m_{\rm 1500}\bigr ] \bigr \}^{2}}
{\sigma^{2} \bigl [ f^{(i)}_{\rm \nu, obs} \bigr ]},
\end{eqnarray}
where $\sigma \bigl [ f^{(i)}_{\rm \nu, obs} \bigr ]$ 
is the photometric and systematic errors in the $i$th bandpass.
The uncertainties in the best-fit parameters correspond to the $1\sigma$ confidence interval, 
$\Delta \chi^{2} < 1.0$. 
With best-fit parameters of $\beta_{\rm 1200-2800}$ and $m_{\rm 1500}$, 
we obtain $f_{\rm \nu, cont}$ from equations (4) and (5). 
The flux $f_{\rm \nu, cont}$ is then converted 
into $f_{\rm \lambda, cont}$ from the relation 
\begin{eqnarray}
f_{\rm \lambda, cont} = \frac{c}{\lambda^{2}} \times f_{\rm \nu, cont}, 
\end{eqnarray}
where $c$ is the speed of light.  
Using equation (10), we derive the continuum flux at 1216 \AA, $f_{\rm cont, 1216}$, 
to obtain EW$_{\rm 0}$(Ly$\alpha$) as 
\begin{eqnarray}
{\rm EW_{\rm 0}(Ly\alpha)} = \frac{F(Ly\alpha)}{f_{\rm cont, 1216}} \times \frac{1}{1+z_{\rm Ly\alpha}}.
\end{eqnarray}
We obtain $M_{\rm UV}$
from the continuum flux at 1500 \AA, $f_{\rm cont, 1500}$, as below: 
\begin{eqnarray}
M_{\rm UV} = m_{1500} -5 {\rm log}(d_{L}/{\rm 10pc}) + 2.5 {\rm log}(1+z_{\rm Ly\alpha}), 
\end{eqnarray}
where $d_{L}$ indicates the luminosity distance corresponding to $z_{\rm Ly\alpha}$.

\begin{figure*}
\centering
\includegraphics[width=12cm]{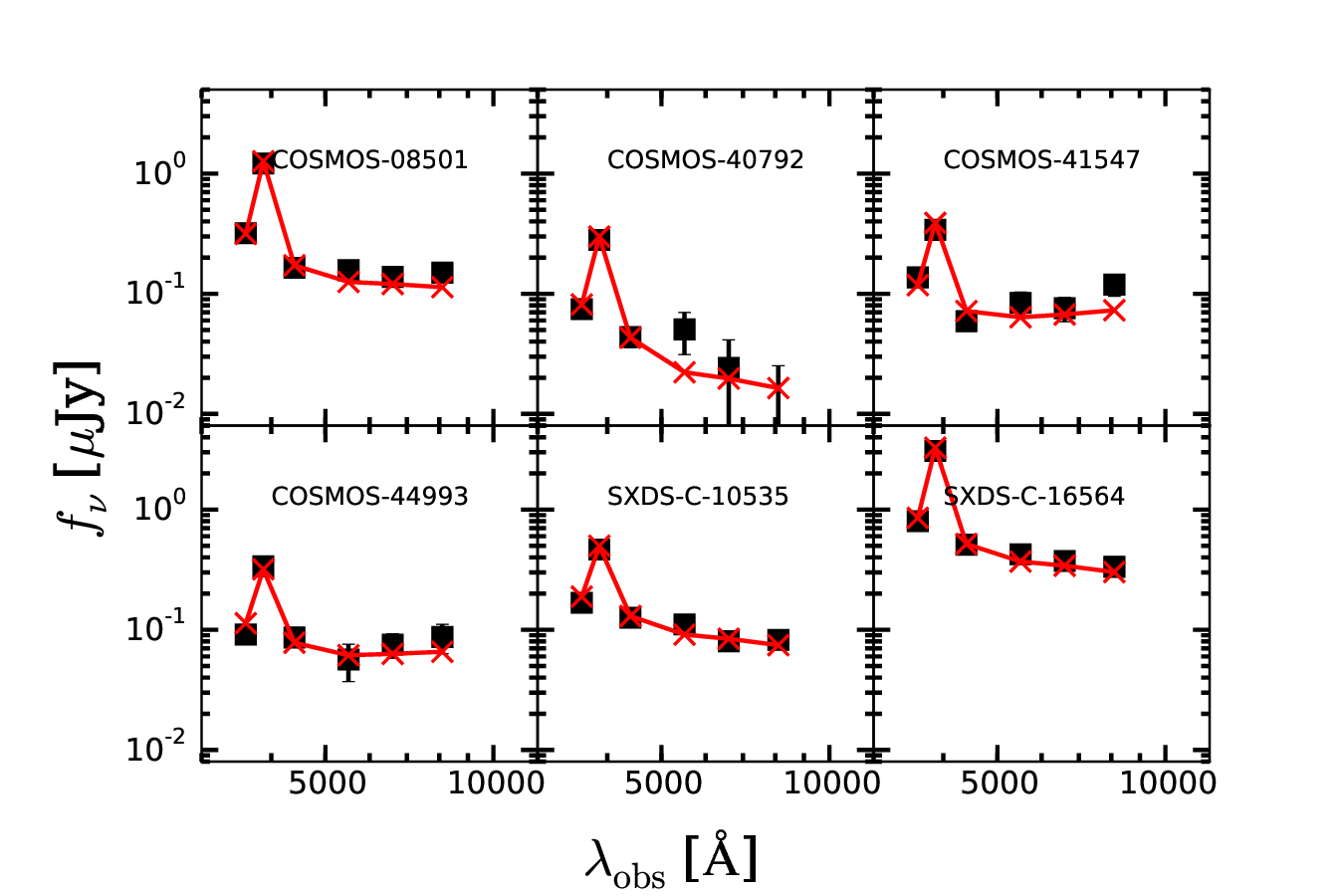}
\caption[]
{
Rest-frame UV SEDs, from ${\it u^{*}}$ to $i'$-band data,  of our LAEs. 
The filled squares denote the photometry used for the fits, 
while the red crosses correspond to
the flux densities at individual passbands 
expected from the best-fit models.
}
\label{fig:ew_beta}
\end{figure*}

\begin{figure}
\centering
\includegraphics[width=\columnwidth]{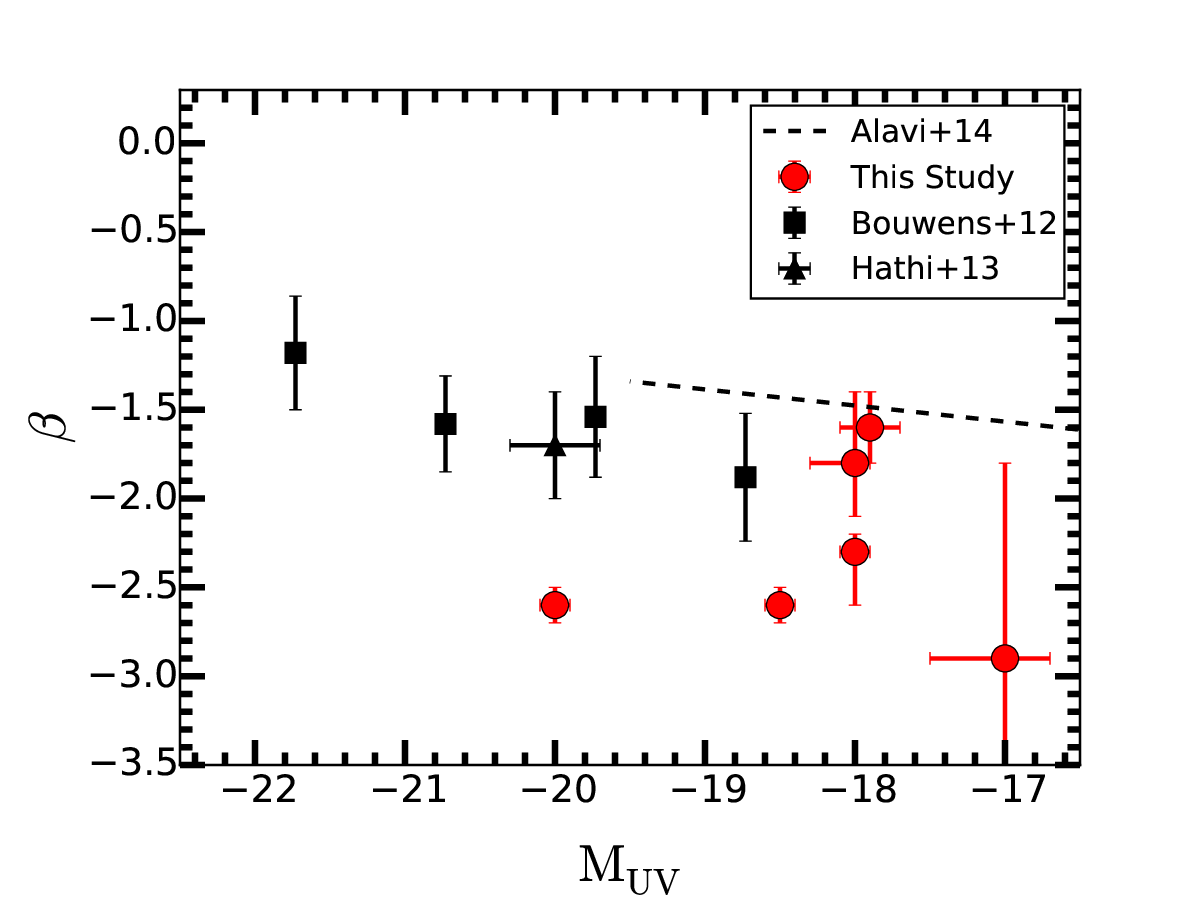}
\caption[]
{
UV spectral slope ($\beta$) as a function of the absolute UV magnitude 
at $1500$\AA\ ($M_{\rm UV}$) at $z\sim2$. 
The red circles are our LAEs, 
where we adopt $\beta_{\rm 1200-2800}$ as the $\beta$ values. 
The dashed line is the best linear fit for $z\sim2$ lensed LBGs \citep{alavi2014}, 
while the black triangle is the average value of $z\sim2$ LBGs \citep{hathi2013}. 
The black squares indicate $z\sim2$ LBGs studied by \cite{bouwens2009}, 
where error bars denote the $1\sigma$ of the distribution at each magnitude bin. 
}
\label{fig:muv_beta}
\end{figure}

Figure \ref{fig:ew_beta} shows the best-fit model spectra. 
As can be seen, our technique reproduces the rest-frame UV SEDs.
The best-fit parameters and their $1\sigma$ uncertainties 
are summarized in Table \ref{tab:EW_beta_MUV}. 
In Table \ref{tab:EW_beta_MUV}, we find that our LAEs have 
large EW$_{\rm 0}$(Ly$\alpha$) values ranging 
from $160$ to $357$ \AA\ with a mean value of $252 \pm 30$ \AA. 
We confirm that LAEs with EW$_{\rm 0}$(Ly$\alpha$) $\gtrsim 200$ \AA\ exist 
by our fitting method with no apriori assumption on UV continuum slopes. 
UV continuum slopes vary from 
$\beta_{\rm 1200-2800} = -1.6$ to $-2.9$ with small mean and median values of 
$-2.3\pm0.2$ and $-2.5\pm0.2$, respectively. 
The median Ly$\alpha$ luminosity of our LAEs is 
$L({\rm Ly\alpha}) = 3.7^{+2.8}_{-2.8} \times10^{42}$ erg s$^{-1}$. 
This is broadly consistent with the characteristic Ly$\alpha$ luminosity 
of $z\sim2$ LAEs obtained by \cite{hayes2010, ciardullo2012} 
and \cite{konno2016}. 
The median UV absolute magnitude of our LAEs is 
$M_{\rm UV} = -18.5$. 
Figure \ref{fig:muv_beta} plots $\beta$ against ${M_{\rm UV}}$ 
for our LAEs and Lyman-break galaxies (LBGs) at $z\sim2$ 
(\citealt{bouwens2009, hathi2013, alavi2014}). 
We note that the error bar of the data points of \cite{bouwens2009} 
indicate the $1\sigma$ of the $\beta$ distribution at each magnitude bin. 
In Figure \ref{fig:muv_beta}, 
our LAEs have $\beta$ values comparable to or smaller 
than the LBGs at a given ${M_{\rm UV}}$ value, 
implying large EW$_{\rm 0}$(Ly$\alpha$) objects have small UV continuum slopes. 
This trend is consistent with previous results (e.g., \citealt{stark2010, hathi2016}).

\begin{figure}
\centering
\includegraphics[width=7cm]{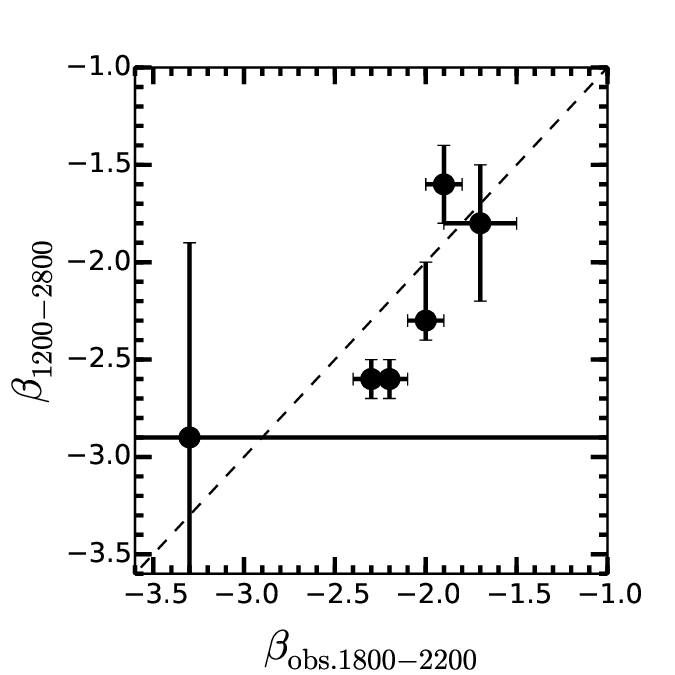}
\caption[]
{
Comparison of the two UV continuum slopes, 
$\beta_{\rm 1200-2800}$ and $\beta_{\rm obs. 1800-2200}$, for our LAEs. 
The dashed line indicates the one-to-one relation. 
}
\label{fig:beta_beta}
\end{figure}

In \S \ref{subsec:age_metal}, we constrain the stellar ages and metallicities of our LAEs 
based on comparisons of the EW$_{\rm 0}$(Ly$\alpha$) and UV continuum slopes  
with stellar evolution models of \cite{schaerer2003} and \cite{raiter2010}. 
Although we have estimated UV continuum slopes at the wavelength range of $1200-2800$ \AA,
\cite{schaerer2003} and \cite{raiter2010} have computed UV continuum slopes 
at the wavelength range of $1800-2200$ \AA. 
Thus, we also calculate UV continuum slopes of our LAEs at the same wavelength range, 
$\beta_{\rm obs. 1800-2200}$, with the following equation:

\begin{eqnarray}
\beta_{\rm obs. 1800-2200}
= - \frac{V-(r'+i')/2}{\rm 2.5 log (\lambda_{\it V}/(\lambda_{\it r'}+\lambda_{\it i'})/2)} - 2,
\end{eqnarray}
where $V$, $r'$, and $i'$ are the magnitudes listed in Table \ref{tab:photometry}, 
while $\lambda_{V}$, $\lambda_{r'}$, and $\lambda_{i'}$ 
correspond to the central wavelengths of each band, $5500$, $6300$, 
and $7700$ \AA, respectively.
We obtain 
$\beta_{\rm obs. 1800-2200}$ $= -2.0\pm0.1$ (COSMOS-08501), 
$-3.3\pm4.0$ (COSMOS-40792), 
$-1.9\pm0.1$ (COSMOS-41547), 
$-1.7\pm0.2$ (COSMOS-44993), 
$-2.3\pm0.1$ (SXDS-C-10535), 
and 
$-2.2\pm0.1$ (SXDS-C-16564).  
Figure \ref{fig:beta_beta} plots $\beta_{\rm 1200-2800}$ against $\beta_{\rm obs.1800-2200}$. 
The data points lie on the one-to-one relation, 
showing that the two UV continuum slopes are consistent with each other.

We note here that
the models of \cite{schaerer2003} and \cite{raiter2010} do not take into account 
dust extinction effects on UV continuum slopes. 
For fair comparisons, 
we derive the intrinsic UV continuum slopes, $\beta_{\rm 1800-2200}$. 
We find that UV continuum slopes increase by 0.5 for $E(B-V)_{\rm *} = 0.1$ 
based on a combination of the empirical relation, $A_{\rm 1600} = 4.43 + 1.99 \beta$ (\citealt{meurer1999}), and Calzetti extinction, $A_{\rm 1600} = k_{\rm 1600} E(B-V)_{\rm *}$
($k_{\rm 1600} = 10$; \citealt{ouchi2004}). 
With $E(B-V)_{\rm *}$ in Table \ref{tab:sed_fit}, 
we obtain 
$\beta_{\rm1800-2200} = -2.4^{+0.2}_{-0.4}$ (COSMOS-08501), 
$-3.3^{+7.9}_{-7.9}$ (COSMOS-40792), 
$-3.1^{+0.3}_{-0.4}$ (COSMOS-41547), 
$-1.9^{+0.6}_{-0.4}$ (COSMOS-44993), 
$-2.3^{+0.1}_{-0.1}$ (SXDS-C-10535), 
and 
$-2.2^{+0.1}_{-0.1}$ (SXDS-C-16564).  
In this calculation, we have adopted $2\sigma$ errors in $\beta_{\rm obs.1800-2200}$ 
to obtain conservative uncertainties in $\beta_{\rm1800-2200}$. 
The mean and median correction factors  
are as small as $-0.3\pm0.2$ and $-0.1\pm0.2$, respectively. 
This is due to the fact that our LAEs have the low median stellar dust extinction value, 
$E(B-V)_{\rm *} = 0.02^{+0.04}_{-0.02}$. 
One might be concerned about the systematic uncertainty of using two different models; 
we have adopted the model of GALAXEV to derive stellar dust extinction 
and the correction factors for UV continuum slopes, 
whereas we use the models of \cite{schaerer2003} and \cite{raiter2010} 
to compare with $\beta_{\rm 1800-2200}$. 
However, the systematic uncertainty is negligibly small 
because our LAEs have small $\beta_{\rm obs.1800-2200}$ values. 
Our conservative uncertainties in $\beta_{\rm1800-2200}$ 
would include these systematic errors.

\begin{table*}
\centering
\caption{Results of Careful Estimates of EW$_{\rm 0}$(Ly$\alpha$) and $\beta$}
\label{tab:EW_beta_MUV}
\begin{tabular}{cccccc}
\hline
{Object} & {$\chi^{2}$} & {EW$_{\rm 0}$(Ly$\alpha$)} 
& {$L({\rm Ly\alpha})$} 
& {$M_{\rm UV}$} &{$\beta_{\rm 1200-2800}$}\\
{} & {} & {(\AA)}  & {(10$^{42}$ erg s$^{-1}$)} & {} & {}\\
{(1)} & {(2)} & {(3)}  & {(4)} & {(5)} & {(6)}\\
\hline
COSMOS-8501 & 8.5 & $284^{+39}_{-16}$ & $8.9^{+0.8}_{-0.5}$ & $-18.0^{+0.1}_{-0.1}$ & $-2.3^{+0.3}_{-0.1}$ \\
\hline
COSMOS-40792 & 2.3 & $357^{+96}_{-114}$ & $2.5^{+0.3}_{-0.4}$ & $-17.0^{+0.5}_{-0.3}$ & $-2.9^{+1.0}_{-1.1}$ \\
\hline
COSMOS-41547 & 13.5 & $303^{+59}_{-46}$ & $3.3^{+0.3}_{-0.4}$ & $-17.9^{+0.2}_{-0.2}$ & $-1.6^{+0.2}_{-0.2}$ \\
\hline
COSMOS-44993 & 2.3 & $215^{+115}_{-22}$ & $2.9^{+0.8}_{-0.4}$ & $-18.0^{+0.3}_{-0.1}$ & $-1.8^{+0.3}_{-0.4}$ \\
\hline
SXDS-C-10535 & 5.8 & $160^{+12}_{-16}$ & $4.1^{+0.4}_{-0.8}$ & $-18.5^{+0.1}_{-0.1}$ & $-2.6^{+0.1}_{-0.1}$ \\
\hline
SXDS-C-16564 & 18.7 & $195^{+7}_{-8}$ & $20.0^{+1.2}_{-1.5}$ & $-20.0^{+0.1}_{-0.1}$ & $-2.6^{+0.1}_{-0.1}$ \\
\hline
\end{tabular}
%
\begin{minipage}{170mm}
\begin{flushleft}
(1) Object ID;
(2) $\chi^{2}$ of the fitting; 
(3) and (4) Rest-frame Ly$\alpha$ EW and Ly$\alpha$ luminosity; 
(5) UV absolute magnitude; 
and 
(6) UV spectral slope at the rest-frame wavelength range of $1800-2200$ \AA.
\end{flushleft}
\end{minipage}
\end{table*}

\subsection{FWHM of Ly$\alpha$ Lines} \label{subsec:fwhm} 

One of the advantages of our LAEs is that they have Ly$\alpha$ detections. 
We examine the FWHM of the Ly$\alpha$ line, FWHM(Ly$\alpha$). 
To derive FWHM(Ly$\alpha$) values, 
we apply a Monte Carlo technique 
exactly the same way as is adopted in \cite{hashimoto2015}. 
Briefly, we measure the $1\sigma$ noise in the Ly$\alpha$ spectrum 
set by the continuum level at wavelengths longer than $1216$ \AA. 
Then we create $10^{3}$ fake spectra 
by perturbing the flux at each wavelength of the true spectrum 
by the measured $1\sigma$ error. 
For each fake spectrum, the wavelength range that encompasses 
half the maximum flux is adopted as the FWHM.  
We adopt the median and standard deviation of the distribution of measurements 
as the median and error values, respectively. 
The measurements corrected for the instrumental resolutions, FWHM$_{\rm int}$(Ly$\alpha$), 
are listed in the column 2 of Table \ref{tab:spec_prop}. 
We do not obtain the FWHM$_{\rm int}$(Ly$\alpha$) of SXDS-C-16564 
because its spectral resolution of the Ly$\alpha$ line, $R=600$, is insufficient 
for a reliable measurement. 
Hereafter, we eliminate SXDS-C-16564 from the sample 
when we discuss the FWHM$_{\rm int}$(Ly$\alpha$) of our LAEs. 
FWHM$_{\rm int}$(Ly$\alpha$) values range from 118 to 310 km s$^{-1}$ 
with a mean value of $212 \pm 32$ km s$^{-1}$.

\begin{figure*}
\centering
\includegraphics[width=17cm]{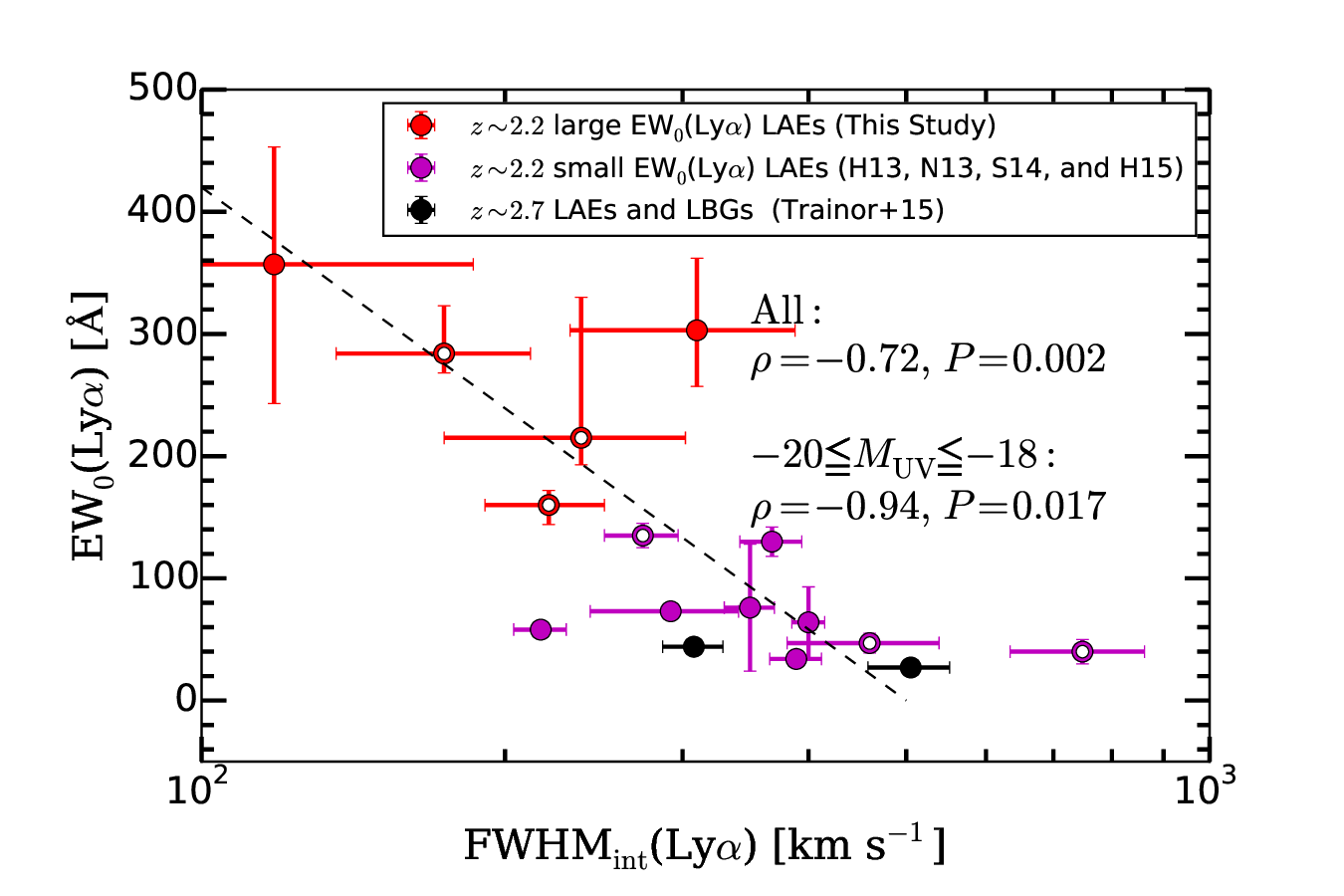}
\caption[]
{
FWHM(Ly$\alpha$) corrected for the instrumental resolution,  FWHM$_{\rm int}$(Ly$\alpha$), 
plotted against  EW$_{\rm 0}$(Ly$\alpha$). Note that the x-axis is in the log-scale. 
The red circles are our large EW$_{\rm 0}$(Ly$\alpha$) LAEs at $z\sim2.2$, 
while 
the magenta circles indicate the small EW$_{\rm 0}$(Ly$\alpha$) LAEs at $z\sim2.2$ 
\citep{hashimoto2013, nakajima2013, shibuya2014b, hashimoto2015}. 
The two black circles show the results for composite spectra of 32 LAEs and 65 LBGs 
at $z\sim2.7$ \citep{trainor2015}. 
For the whole sample, 
the Spearman rank correlation coefficient for the relation is $\rho = -0.72$, 
while the probability satisfying the null hypothesis is $P=0.002$.
The dashed line is the linear fit to the data points. 
The six open circles indicate 
the LAEs with similar $M_{\rm UV}$ values, $-20 \leqq M_{\rm UV} \leqq -18$, respectively. 
For the six LAEs, the Spearman rank correlation test gives $\rho = -0.94$ and $P=0.017$. 
}
\label{fig:fwhm_ew}
\end{figure*}

For comparisons, we also measure FWHM$_{\rm int}$(Ly$\alpha$) values of 
nine $z\sim2$ LAEs with small EW$_{\rm 0}$(Ly$\alpha$) values in the literature 
(\citealt{hashimoto2013, nakajima2013, shibuya2014b, hashimoto2015}). 
Among the LAEs studied in these studies, 
we do not use COSMOS-30679 whose Ly$\alpha$ emission is contaminated 
by a cosmic ray (\citealt{hashimoto2013}). 
Hereafter, we refer this sample as ``small EW$_{\rm 0}$(Ly$\alpha$) LAEs''. 
The mean EW$_{\rm 0}$(Ly$\alpha$) is $65\pm10$ \AA, 
while the mean FWHM$_{\rm int}$(Ly$\alpha$) is calculated to be $389\pm51$ km s$^{-1}$. 
Table \ref{tab:prop_low} summarizes 
the EW$_{\rm 0}$(Ly$\alpha$) and FWHM$_{\rm int}$(Ly$\alpha$) values 
of the small EW$_{\rm 0}$(Ly$\alpha$) LAEs. 
\footnote{
We note here that COSMOS-43982 has a signature of an AGN activity 
(\citealt{nakajima2013, shibuya2014b, hashimoto2015}). 
We have confirmed that  our discussion remains unchanged 
whether or not we include this object into the small EW$_{\rm 0}$(Ly$\alpha$) LAEs. 
}
In addition, \cite{trainor2015} have also investigated Ly$\alpha$ profiles 
of LAEs at $z\sim2.7$. 
For the composite spectrum of 32 LAEs that have both Ly$\alpha$ and nebular line detections, 
the typical EW$_{\rm 0}$(Ly$\alpha$) value is $44$ \AA, 
while the mean FWHM$_{\rm int}$(Ly$\alpha$) value is $309\pm22$ km s$^{-1}$.
Using a sample of the large EW$_{\rm 0}$(Ly$\alpha$) LAEs, 
the small EW$_{\rm 0}$(Ly$\alpha$) LAEs, 
and the LAEs and LBGs in \cite{trainor2015}, 
we plot EW$_{\rm 0}$(Ly$\alpha$) 
as a function of FWHM$_{\rm int}$(Ly$\alpha$) in Figure \ref{fig:fwhm_ew}. 
In this figure, the data points of \cite{trainor2015} 
cover the small EW$_{\rm 0}$(Ly$\alpha$) range 
complementary to our LAE results. 
We carry out the Spearman rank correlation test to evaluate the significance of a correlation. 
The rank correlation coefficient is $\rho = -0.72$, 
while the probability satisfying the null hypothesis is $P=0.002$. 
The result indicates that FWHM$_{\rm int}$(Ly$\alpha$) anti-correlates with EW$_{\rm 0}$(Ly$\alpha$). 
We also carry out the Spearman rank correlation test 
for objects with similar $M_{\rm UV}$ values. 
For six LAEs satisfying $-20 \leqq M_{\rm UV} \leqq -18$ 
(open circles in Figure \ref{fig:fwhm_ew}), 
we obtain $\rho = -0.94$ and $P=0.016$. 
The result confirms that the anti-correlation is not due to the selection effect 
in $M_{\rm UV}$. 
Although \cite{tapken2007} have claimed a qualitatively similar anti-correlation 
between EW$_{\rm 0}$(Ly$\alpha$) and FWHM$_{\rm int}$(Ly$\alpha$) 
for their small EW$_{\rm 0}$(Ly$\alpha$) LAEs at a high-$z$ range of $z\sim2.7-4.5$, 
no correlation test has been carried out. 
In our study, we have identified for the first time the anti-correlation 
based on a statistical test. 
Moreover, we have found the anti-correlation at the range of 
EW$_{\rm 0}$(Ly$\alpha$) $\gtrsim 200$ \AA.

Several other studies have also studied FWHM$_{\rm int}$(Ly$\alpha$) values of LAEs 
at a high-$z$ range of $z\sim3-7$. 
\cite{tapken2007} have investigated 
EW$_{\rm 0}$(Ly$\alpha$) and FWHM$_{\rm int}$(Ly$\alpha$) values 
of individual LAEs at $z\sim2.7-4.5$. 
In this study, 
the mean EW$_{\rm 0}$(Ly$\alpha$) is $47\pm13$ \AA, 
while the mean FWHM$_{\rm int}$(Ly$\alpha$) is $472\pm53$ km s$^{-1}$. 
These values are consistent with those of the small EW$_{\rm 0}$(Ly$\alpha$) LAEs 
(Table \ref{tab:prop_low}). 
At $z=5.7$ and $6.6$, 
\cite{ouchi2010} have measured FWHM$_{\rm int}$(Ly$\alpha$) values 
of composite spectra of LAEs. 
The sample of \cite{ouchi2010} do not include large EW$_{\rm 0}$(Ly$\alpha$) LAEs. 
Nevertheless, the mean FWHM$_{\rm int}$(Ly$\alpha$) values are 
$265\pm37$ km s$^{-1}$ and  $270\pm16$ km s$^{-1}$
for $z=5.7$ and $6.6$, respectively, 
smaller than those of the small EW$_{\rm 0}$(Ly$\alpha$) LAEs (Table \ref{tab:prop_low}). 
This would be due to strong Ly$\alpha$ scattering in the IGM at $z\sim6-7$ 
compared to that at $z\sim2$: 
the IGM scattering significantly narrows the blue part of Ly$\alpha$ profile at $z\sim6-7$
(\citealt{laursen2011}).

\begin{table*}
\centering
\caption{Summary of Spectroscopic Properties of our Large EW$_{\rm 0}$(Ly$\alpha$) LAEs}
\label{tab:spec_prop}
\begin{tabular}{ccccc}
\hline
{Object}  & {FWHM$_{\rm int}$(Ly$\alpha$)} 
& {$3\sigma$ $f_{\rm He{\sc II}}/f_{\rm Ly\alpha}$} & {3$\sigma$ EW$_{\rm 0}$(He{\sc II})} \\
{} &  {(km s$^{-1}$)} & {(\AA)} \\
{(1)} & {(2)} & {(3)} & {(4)} \\
\hline
COSMOS-08501 &  $174\pm38$  & $-$ & $-$\\
\hline
COSMOS-40792 &  $118\pm68$  & $0.11^{+0.01}_{-0.02}$ & $91^{+38}_{-27}$\\
\hline
COSMOS-41547 &  $310\pm78$ & $0.10^{+0.01}_{-0.01}$ & $40^{+4}_{-3}$ \\
\hline
COSMOS-44993 &  $238\pm64$  & $0.12^{+0.03}_{-0.02}$ & $41^{+5}_{-4}$\\
\hline
SXDS1-10535 & $221\pm30$   & $0.08^{+0.01}_{-0.02}$ & $18^{+2}_{-2}$\\
\hline
SXDS1-16564 &  $-$  & $0.02^{+0.01}_{-0.01}$ & $7^{+2}_{-2}$\\
\hline
\end{tabular}
%
\begin{minipage}{100mm}
\begin{flushleft}
The symbol g-h indicates we have no measurement. 
(1) Object ID;
(2) FWHMs of the Ly$\alpha$ lines corrected for the instrumental resolutions; 
(3) 3$\sigma$ upper limits of the flux ratio of He{\sc ii} and Ly$\alpha$; 
and 
(4) 3$\sigma$ upper  limits of the rest-frame He{\sc ii} EW. 
\end{flushleft}
\end{minipage}
\end{table*}

\begin{table*}
\centering
\caption{Properties of small EW$_{\rm 0}$(Ly$\alpha$) LAEs}
\label{tab:prop_low}
\begin{tabular}{cccccc}
\hline
{Object}  &  {EW$_{\rm 0}$(Ly$\alpha$)} & {FWHM$_{\rm int}$(Ly$\alpha$)}  & {Source$^{a}$}  \\
{} & {(\AA)} & {(km s$^{-1}$)}  & {} \\
{(1)} & {(2)} & {(3)} & {(4)} \\
\hline
CDFS-3865 & $64\pm29$ & $400\pm15$  & H13, N13, H15\\
\hline
CDFS-6482 & $76\pm52$ & $350\pm20$  & H13, N13, H15\\
\hline
COSMOS-13636 & $73\pm5$ & $292\pm49$  & H13, N13, H15\\
\hline
COSMOS-43982$^{b}$ & $130\pm12$ & $368\pm26$ & H13, N13, H15\\
\hline
COSMOS-08357 & $47\pm8$ & $460\pm79$  & S14, H15\\
\hline
COSMOS-12805 & $34\pm6$ & $389\pm23$  & S14, H15\\
\hline
COSMOS-13138 & $40\pm10$ & $748\pm114$  & S14, H15\\
\hline
SXDS-10600 & $58\pm3$ & $217\pm13$  & S14, H15\\
\hline
SXDS-10942 & $135\pm10$ & $274\pm23$   & S14, H15\\
\hline
\end{tabular}
%
\begin{minipage}{100mm}
\begin{flushleft}
(1) Object ID;
(2) Rest-frame Ly$\alpha$ EWs; 
(3) FWHMs of the Ly$\alpha$ lines corrected for the instrumental resolutions; 
and 
(4) Source of the information 

$^a$
H13: \cite{hashimoto2013}; N13: \cite{nakajima2013}; S14: \cite{shibuya2014b}; H15: \cite{hashimoto2015}

$^b$
AGN-like object
\end{flushleft}
\end{minipage}
\end{table*}

\subsection{Upper Limits on the Flux Ratio of He{\sc ii}/Ly$\alpha$ and EW$_{\rm 0}$(He{\sc ii})} \label{subsec:upper_limit_helium}

We derive $3 \sigma$ upper limits of the flux ratio, $f_{\rm He II}$/$f_{\rm Ly\alpha}$, 
where $f_{\rm He II}$ and $f_{\rm Ly\alpha}$ are the He{\sc ii} and Ly$\alpha$ fluxes, 
respectively.
We do not derive the flux ratio for COSMOS-08501 
whose FUV data have been obtained with MagE. 
This is because the flux calibration of echelle spectra is often inaccurate (\citealt{willmarth1994}). 
Following the procedure in \cite{kashikawa2012}, 
we obtain the $3 \sigma$ upper limits of the He{\sc ii} fluxes. 
These He{\sc ii} fluxes are given  from the wavelength ranges of 8.8 (4.8) \AA\ 
for the IMACS (LRIS) spectra 
under the assumptions that the He{\sc ii} lines are not resolved. 
The derived $3\sigma$ upper limits are  
$f_{\rm He II}$/$f_{\rm Ly\alpha} = 0.11^{+0.01}_{-0.02}$ (COSMOS-40792), 
$0.10^{+0.01}_{-0.01}$ (COSMOS-41547), 
$0.12^{+0.03}_{-0.02}$ (COSMOS-44993), 
$0.08^{+0.01}_{-0.02}$ (SXDS-C-10535), 
and 
$0.02^{+0.01}_{-0.01}$ (SXDS-C-16564) 
(the column 3 of Table \ref{tab:spec_prop}).
These $3\sigma$ upper limits are stronger than 
the $2\sigma$ upper limit of $f_{\rm He II}$/$f_{\rm Ly\alpha} = 0.23$ 
derived for a strong LAE at $z=6.3$ (\citealt{nagao2005}). 
Moreover, 
these $3\sigma$ upper limits are comparable to the $3\sigma$ upper limits of 
$f_{\rm He II}$/$f_{\rm Ly\alpha} \sim 0.02-0.06$ 
obtained for LAEs at $z\sim3.1-3.7$ (\citealt{ouchi2008}) 
and at $z=6.5$ (\citealt{kashikawa2012}). 
Recently, \cite{sobral2015} have reported the He{\sc ii} line detection 
from a strong LAE at $z=6.6$, CR7, at the significance level of $6\sigma$. 
In this study, 
the rest-frame EW, EW$_{\rm 0}$(He{\sc ii}), 
is measured to be $\sim 80$ \AA\
(see also \citealt{bowler2016} 
who have obtained EW$_{\rm 0}$(He{\sc ii}) $=40\pm30$ \AA\ 
with deep near-infrared photometric data).
The measured flux ratio of CR7 is $f_{\rm He II}$/$f_{\rm Ly\alpha} = 0.23\pm0.10$. 

We calculate the fraction of large EW$_{\rm 0}$(Ly$\alpha$) LAEs 
with He{\sc ii} detections among large EW$_{\rm 0}$(Ly$\alpha$) LAEs, 
combining our results with those in the literature. 
There are nine LAEs that satisfy 
EW$_{\rm 0}$(Ly$\alpha$) $\gtrsim$ 130 \AA. 
These LAEs include 
five, one, one, and two objects 
from 
this study, 
\cite{nagao2005}, 
\cite{kashikawa2012}, 
and \cite{sobral2015}, respectively. 
We thus estimate the fraction to be $\sim$ $10\%$ (1/9).

We also examine $3\sigma$ upper limits of the EW$_{\rm 0}$(He{\sc ii}). 
To do so, we derive the continuum flux at 1640 \AA\ from photometric data 
with fitting results (\S \ref{subsec:estimate_ew_beta}). 
These estimates give us $3\sigma$ limits of 
EW$_{\rm 0}$(He{\sc ii}) $\leqq$ $91^{+38}_{-27}$ \AA\ (COSMOS-40792), 
$40^{+4}_{-3}$ \AA\ (COSMOS-41547),
$41^{+5}_{-4}$ \AA\ (COSMOS-44993),
$18^{+2}_{-2}$ \AA\ (SXDS-C-10535),
and 
$7^{+2}_{-2}$ \AA\ (SXDS-C-16564) (the column 4 of Table \ref{tab:spec_prop}). 
We use the $3\sigma$ upper limits of the EW$_{\rm 0}$(He{\sc ii}) 
to place constraints on the stellar ages and metallicities of our LAEs (\S\ref{subsec:age_metal}).

\subsection{Coarse Estimates of the Ly$\alpha$ Escape Fraction} \label{subsubsec:fesc}

The Ly$\alpha$ escape fraction, $f^{\rm Ly\alpha}_{\rm esc}$, 
is defined as the ratio of the observed Ly$\alpha$ flux 
to the intrinsic Ly$\alpha$ flux produced in a galaxy. 
This quantity is mainly determined by 
a neutral hydrogen column density, $N_{\rm HI}$, or a dust content in the ISM.
If the ISM has a low $N_{\rm HI}$ value or a low dust content, 
a high $f^{\rm Ly\alpha}_{\rm esc}$ value is expected 
because Ly$\alpha$ photons are less scattered and absorbed by dust grains (e.g., \citealt{atek2009, hayes2011, cassata2015})
\footnote{
The outflowing ISM also facilitates the Ly$\alpha$ escape 
due to the reduced number of scattering (e.g., \citealt{kunth1998, atek2008, rivera-thorsen2015}). 
}

Many previous studies have estimated 
Ly$\alpha$ escape fractions 
on the assumptions of  Case B, the Salpeter IMF, and the Calzetti's dust extinction law. 
These assumptions would increase systematic uncertainties in the estimates 
of the Ly$\alpha$ escape fractions. 
Nevertheless, in order to compare Ly$\alpha$ escape fractions 
of our LAEs with those in the literature, 
we obtain Ly$\alpha$ escape fractions conventionally as 

\begin{eqnarray}
f^{\rm Ly\alpha}_{\rm esc} = \frac{L_{\rm obs}({\rm Ly}\alpha)}{L_{\rm int}({\rm Ly}\alpha)}, 
\end{eqnarray}

where subscripts ``int'' and ``obs'' refer to intrinsic and observed quantities, respectively. 
We infer $L_{\rm int}({\rm Ly}\alpha)$ from the SFRs in Table \ref{tab:sed_fit} 
using 
$L_{\rm int}({\rm Ly}\alpha)$ [erg s$^{-1}$]  $ = 1.1 \times 10^{42}$ SFR [$M_{\rm \odot}$ yr$^{-1}$] 
(\citealt{kennicutt1998}) on the assumption of Case B. 
For COSMOS-08501 that has the H$\alpha$ detection, 
we quote the $f^{\rm Ly\alpha}_{\rm esc}$ value estimated from
the extinction-corrected  H$\alpha$ luminosities calculated by \cite{nakajima2013}. 
We have obtained $f^{\rm Ly\alpha}_{\rm esc}$ = 
$1.21^{+0.31}_{-0.38}$ (COSMOS-08501), 
$5.33^{+2045}_{-0.91}$ (COSMOS-40792), 
$0.16^{+0.26}_{-0.11}$ (COSMOS-41547), 
$1.55^{+294}_{-0.85}$ (COSMOS-44993), 
$0.59^{+18}_{-0.45}$ (SXDS-C-10535), 
and 
$0.68^{+0.04}_{-0.05}$ (SXDS-C-16564). 
For the three objects that have relatively small errors, 
COSMOS-08501, COSMOS-41547, and SXDS-C-16564, 
the mean and median Ly$\alpha$ escape fractions are 
$f^{\rm Ly\alpha}_{\rm esc} = 0.68\pm0.30$ and $0.68\pm0.30$, respectively. 
These values are much higher than 
the average Ly$\alpha$ escape fraction of $z\sim2$ galaxies, 
$f^{\rm Ly\alpha}_{\rm esc}$ $\sim 2-5 \%$ 
(\citealt{hayes2010, steidel2011, ciardullo2014, oteo2015, matthee2016}), 
and even higher than the average value of $z\sim2$ LAEs, $f^{\rm Ly\alpha}_{\rm esc}$ $\sim10-37 \%$ 
(\citealt{steidel2011, nakajima2012, kusakabe2015, trainor2015, erb2016}).

\begin{figure*}
\centering
\includegraphics[width=16cm]{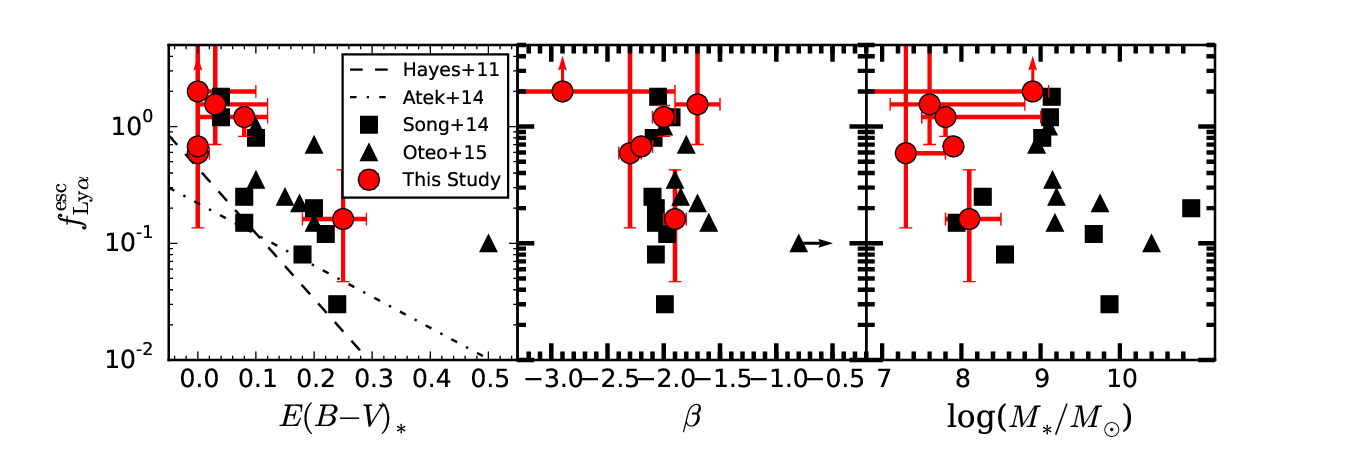}
\caption[]
{
$f^{\rm Ly\alpha}_{\rm esc}$ plotted against $E(B-V)_{\rm *}$, $\beta$, and $M_{\rm *}$. 
The red circles denote our LAEs. 
The black squares are nine LAEs at $z\sim2$ with Ly$\alpha$ and H$\alpha$ detections 
\citep{song2014}, whereas the black triangles show seven LAEs at $z\sim2$ 
with Ly$\alpha$ and H$\alpha$ \citep{oteo2015}. 
In the left panel, the dashed and dot-dashed lines indicate 
the relation between $f^{\rm Ly\alpha}_{\rm esc}$ and $E(B-V)_{\rm *}$ 
for $z\sim0-1$ \citep{atek2014} and $z\sim2-3$ galaxies \citep{hayes2011}, 
respectively. 
In the middle panel, we adopt $\beta_{\rm 1800-2200}$ for our LAEs.
}
\label{fig:fesc}
\end{figure*}

Figure \ref{fig:fesc} plots $f^{\rm Ly\alpha}_{\rm esc}$ against 
$E(B-V)_{*}$, $\beta$, and $M_{\rm *}$. 
We also plot the data points of $z\sim2$ LAEs studies by \cite{song2014} and \cite{oteo2015} 
with both Ly$\alpha$ and H$\alpha$ detections. 
In these studies, 
Ly$\alpha$ escape fractions 
have been estimated with H$\alpha$ luminosities. 
Although no individual measurements of UV continuum slopes are given in \cite{song2014}, 
we calculate $\beta_{\rm 1800-2200}$ values of the LAEs 
with equation (13) using the $V, r,$ and $i-$band photometry listed in Table 3 
of \cite{song2014}. For the consistency, we adopt $\beta_{\rm 1800-2200}$ for our LAEs. 
\cite{oteo2015} have shown that $f^{\rm Ly\alpha}_{\rm esc}$ anti-correlates with 
$E(B-V)_{*}$, $\beta$, and $M_{\rm*}$. 
The result of \cite{oteo2015} indicates that 
Ly$\alpha$ photons preferentially escape 
from low-mass and low dust content galaxies. 
With the median values of 
$E(B-V)_{*}$  = $0.02^{+0.04}_{-0.02}$, 
$\beta_{1800-2200}$ = $-2.2\pm0.2$,
and 
$M_{\rm*}$ =  $7.9^{+4.6}_{-2.9}\times10^{7}$ $M_{\rm \odot}$, 
our LAEs can be regarded as the extreme cases in these trends.

\section{Discussion} \label{sec:discussion}

\subsection{Mode of Star Formation} \label{subsec:sf_mode}

There is a relatively tight relation between SFRs and stellar masses of galaxies 
called the star formation main sequence (SFMS) (e.g., \citealt{daddi2007, rodighiero2011, speagle2014}). 
Galaxies lie on the SFMS are thought to be in a long-term constant star-formation mode, 
while those lie above the SFMS are forming stars in a rapid starburst mode (\citealt{rodighiero2011}).  
We note here that the star-formation mode is different from the SFH (see \S \ref{subsec:sed_fit}). 
As explained, star-formation mode refers to the position of a galaxy 
in the relation between SFRs and stellar masses. 
In contrast, SFHs express the functional forms of SFRs, 
e.g., $e^{-t/\tau}$ for the exponentially declining SFH, 
where $t$ and $\tau$ indicate the age and the typical timescale, respectively. 
The burst SFH indicates the declining SFH with $\tau < 100$ Myr (e.g., \citealt{hathi2016}). 
In the case of the constant SFH, the SFR is constant over time. 
With these in mind, 
we investigate the mode of star-formation of our LAEs 
with SFRs and stellar masses in \S \ref{subsec:sed_fit}. 

\begin{figure*}
\centering
\includegraphics[width=15cm]{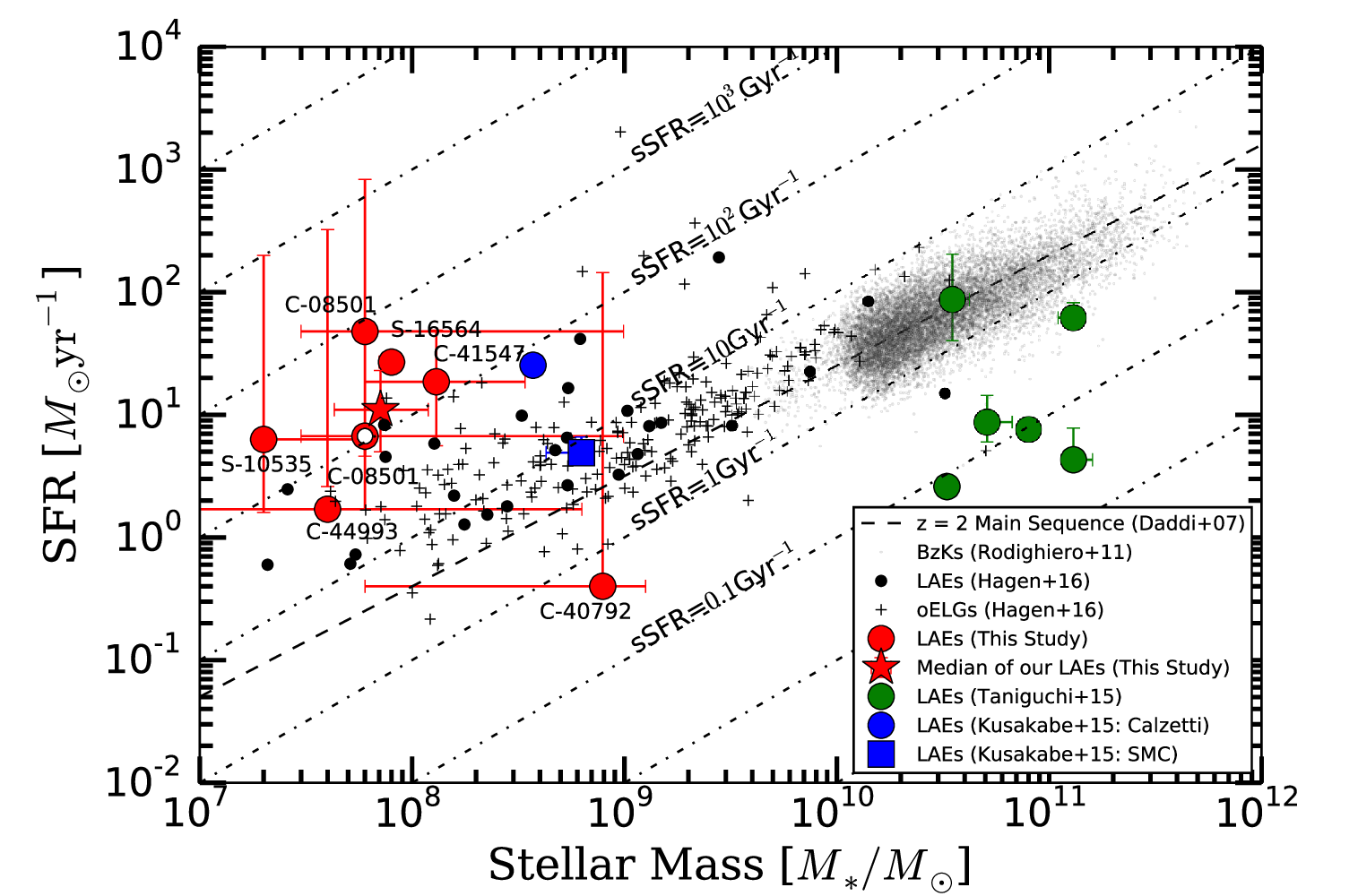}
\caption[]
{
SFRs plotted against $M_{\rm *}$ for our large EW$_{\rm 0}$(Ly$\alpha$) LAEs 
and objects at $z\sim2-3$ in the literature, 
where dot-dashed lines represent specific SFRs (sSFR $\equiv$ SFR/$M_{\rm *}$). 
The red filled circles show our six large EW$_{\rm 0}$(Ly$\alpha$) LAEs  
with a median stellar mass of $M_{\rm *} = 7.1^{+4.8}_{-2.8}\times 10^{7} M_{\odot}$. 
The red open circle is COSMOS-08501 whose SFR estimated 
from the extinction-corrected H$\alpha$ luminosities (\citealt{nakajima2013}).
The red star means the median of the six LAEs. 
The gray dots indicate $z\sim2$ BzK galaxies \citep{rodighiero2011}, 
while the dashed line shows the SFMS at z = 2 \citep{daddi2007}. 
The black circles and crosses are luminous (L(Ly$\alpha$)$> 10^{43}$ erg s$^{-1}$) LAEs 
and optical emission line (e.g., [{\sc Oii}], H$\beta$, and [{\sc Oiii}]) galaxies (oELGs) 
at $z\sim2$ studied by \cite{hagen2016}, respectively. 
The green circles indicate $z\sim3$ LAEs with large EW$_{\rm 0}$(Ly$\alpha$) values 
{\it and} evolved stellar populations \citep{taniguchi2015}. 
The blue circle and square denote the results of the stacking of 214 $z\sim2$ LAEs 
on the assumption of the Calzetti's curve and the SMC attenuation curve, respectively \citep{kusakabe2015}. 
}
\label{fig:mass_sfr}
\end{figure*}

Figure \ref{fig:mass_sfr} plots SFRs against stellar masses for our LAEs. 
Figure \ref{fig:mass_sfr} also includes the data points of 
LAEs in the literature (\citealt{kusakabe2015, taniguchi2015, hagen2016}), 
BzK galaxies (\citealt{rodighiero2011}), 
and optical emission line galaxies (\citealt{hagen2016}) at $z\sim2-3$. 
For COSMOS-08501, we also plot its SFR estimated 
from 
the extinction-corrected H$\alpha$ luminosities calculated by \cite{nakajima2013}.
In Figure \ref{fig:mass_sfr}, 
the median of the six large EW$_{\rm 0}$(Ly$\alpha$) LAEs is shown as the red star. 
The median data point indicates that 
our LAEs are typically lie above the lower-mass extrapolation of the $z \sim 2$ SFMS 
(\citealt{daddi2007, speagle2014}). 
The specific SFRs (sSFR $\equiv$ SFR/$M_{\rm *}$) of our LAEs 
are mostly in the range of sSFR $= 10-1000$ Gyr$^{-1}$ 
with a median value of $\sim100$ Gyr$^{-1}$. 
The median sSFR of our LAEs is higher than those of LAEs and oELGs 
at $z\sim2$ in \cite{hagen2016}, $\sim10$ Gyr$^{-1}$.

Before interpreting the result, we note that stellar masses and SFRs in this study 
are derived from SED fitting on the assumption of the constant SFH. 
Thus, we need to check if our LAEs have high sSFRs on the assumption of 
other SFHs. \cite{schaerer2013} have examined how physical quantities depend on 
the choice of SFHs. 
This study includes exponentially declining, 
exponentially rising, and constant SFHs.
As can be seen from Figures 4 and 7 in \cite{schaerer2013}, 
stellar masses (SFRs) are the largest (smallest) for the constant SFH case 
among the various SFH cases. 
This means that the true sSFRs of our LAEs could be larger 
than what we have obtained.
Therefore, our LAEs have high sSFRs regardless of the choice of SFHs.

A straightforward interpretation of the offset 
toward the high sSFR  
is that our large EW$_{\rm 0}$(Ly$\alpha$) LAEs are in the burst star-formation mode. 
As discussed in detail by \cite{hagen2016}, 
the offset can be also due to (i) a possible change 
in the slope of the SFMS at the low-mass range, 
(ii) errors in the estimates of SFRs and stellar masses, 
or due to (iii) the selection bias against objects with high sSFRs at the low-mass range. 
As to the second point, \cite{kusakabe2015} have shown that 
typical LAEs favor the Small Magellanic Cloud (SMC) attenuation curve (\citealt{pettini1998}) 
rather than the Calzetti's curve (\citealt{calzetti2000}). 
\cite{kusakabe2015} have demonstrated that SFRs are roughly ten times overestimated 
if one uses the Calzetti's curve (blue symbols in Figure \ref{fig:mass_sfr}). 
However, we stress that our estimates of SFRs and stellar masses remain unchanged 
regardless of the extinction curve because our LAEs have small UV continuum slopes. 
Therefore, the second scenario is unlikely for our LAEs. 
As to the third point, \cite{shimakawa2016} 
have investigated SFRs and stellar masses of LAEs 
with $M_{\rm *} > 10^{8} M_{\rm \odot}$ at $z\sim2.5$. 
In contrast to our results and those in \cite{hagen2016}, 
LAEs in \cite{shimakawa2016} follow the SFMS. 
Thus, it is possible that the high sSFRs of our LAEs 
are simply due to the selection bias. 
A large and uniform sample of galaxies with $M_{\rm *}=10^{7-8}$ 
is needed for a definitive conclusion. 

Recently, \cite{taniguchi2015} have reported six rare LAEs at $z\sim3$ 
that have large EW$_{\rm 0}$(Ly$\alpha$) values {\it and} evolved stellar populations. 
Their EW$_{\rm 0}$(Ly$\alpha$) values range from 
$107$ to 306 \AA\ with a mean value of $188\pm30$ \AA. 
\cite{taniguchi2015} have found that these LAEs lie {\it below} the SFMS, 
suggesting that these LAEs are ceasing star forming activities. 
Based on the fact that our LAEs and those in \cite{taniguchi2015} 
have similar EW$_{\rm 0}$(Ly$\alpha$) values,  
the EW$_{\rm 0}$(Ly$\alpha$) value is not necessarily 
a good indicator of the mode of star-formation.

\subsection{
Interpretations of the Small FWHM$_{\rm int}$(Ly$\alpha$) 
in Large EW$_{\rm 0}$(Ly$\alpha$) LAEs } \label{subsubsec:fwhm_ew}

In Figure \ref{fig:fwhm_ew}, we have 
demonstrated that there is an anti-correlation 
between EW$_{\rm 0}$(Ly$\alpha$) and FWHM$_{\rm int}$(Ly$\alpha$). 
In this relation, 
our large EW$_{\rm 0}$(Ly$\alpha$) LAEs have 
small FWHM$_{\rm int}$(Ly$\alpha$) values. 
We give three interpretations of the small FWHM$_{\rm int}$(Ly$\alpha$) 
in our LAEs
\footnote{
\cite{zheng2014} have performed Ly$\alpha$ radiative transfer calculations 
with an anisotropic {\sc Hi} gas density. 
As can be seen from Figure 4 in \cite{zheng2014}, 
for a given $N_{\rm HI}$, the anisotropic {\sc Hi} gas density results in the anti-correlation 
between EW$_{\rm 0}$(Ly$\alpha$) and FWHM$_{\rm int}$(Ly$\alpha$).
Thus, our results might simply indicate that the anisotropic {\sc Hi} gas density 
is at work in LAEs. 
}.

First, assuming uniform expanding shell models, 
\cite{verhamme2015} have theoretically shown that 
the small FWHM$_{\rm int}$(Ly$\alpha$) value is expected 
in the case of a low $N_{\rm HI}$ value in the ISM (see Figure 1 of \citealt{verhamme2015}). 
If the physical picture of the theoretical study is true, 
the small FWHM$_{\rm int}$(Ly$\alpha$) of our large EW$_{\rm 0}$(Ly$\alpha$) LAEs 
suggest that 
our LAEs would have low $N_{\rm HI}$ values in the ISM. 

Second, \cite{gronke2016} have performed Ly$\alpha$ radiative transfer calculations 
of multiphase ISM models.   
The result shows that narrow Ly$\alpha$ profiles can be reproduced by two cases, 
one of which is on the condition that a galaxy has a low covering fraction of the neutral gas
\footnote
{
Another case is the low temperature and number density of the {\sc Hi} gas 
in the inter-clump medium of the multiphase ISM. 
}. 
Thus, our large EW$_{\rm 0}$(Ly$\alpha$) LAEs may have 
lower covering fractions of the neutral gas 
than small EW$_{\rm 0}$(Ly$\alpha$) objects. 
Indeed, based on the analysis of the EW of low-ionization metal absorption lines, 
several studies have observationally shown that 
the neutral-gas covering fraction is low for galaxies with strong Ly$\alpha$ emission 
(e.g., \citealt{jones2013, shibuya2014b, trainor2015}) 

Finally, 
on the assumption that the FWHM$_{\rm int}$(Ly$\alpha$) value 
is determined by a dynamical mass to the first order, 
the small FWHM$_{\rm int}$(Ly$\alpha$) values of our large EW$_{\rm 0}$(Ly$\alpha$) LAEs 
imply that our LAEs would have low dynamical masses 
compared to small EW$_{\rm 0}$(Ly$\alpha$) objects. 
Although we admit that the FWHM$_{\rm int}$(Ly$\alpha$) value 
is dominantly determined by radiative transfer effects 
rather than dynamical masses, 
there is an observational result that may support this interpretation. 
\cite{hashimoto2015} and \cite{trainor2015} have found that 
EW$_{\rm 0}$(Ly$\alpha$) anti-correlates with 
the FWHM value of nebular emission lines (e.g., H$\alpha$, [{\sc Oiii}]), FWHM(neb). 
Since the FWHM(neb) value should correlate with the dynamical mass 
(e.g., \citealt{erb2006b, erb2014}), 
the anti-correlation between EW$_{\rm 0}$(Ly$\alpha$) and FWHM(neb) 
means that large EW$_{\rm 0}$(Ly$\alpha$) LAEs have low dynamical masses. 
Among the large EW$_{\rm 0}$(Ly$\alpha$) LAEs, 
COSMOS-8501 has the H$\alpha$ detection. 
However, only an upper limit of the FWHM of the H$\alpha$ line is derived 
because the line is not resolved (FWHM(neb) $<200$ km s$^{-1}$; \citealt{hashimoto2015}). 
This prevents us from obtaining a definitive conclusion 
on which of the three interpretations are likely for our LAEs.

\subsection{Constraints on Stellar Ages and Metallicities} \label{subsec:age_metal}

We place constraints on the stellar ages and metallicities of our LAEs 
by comparisons of our observational constrains of $\beta$, 
EW$_{\rm 0}$(He{\sc ii}), and EW$_{\rm 0}$(Ly$\alpha$) 
with the stellar evolution models of \cite{schaerer2003} and \cite{raiter2010}. 

\subsubsection{Stellar Evolutionary Models} \label{subsubsec:age_metal_model}

\cite{schaerer2003} and the extended work of \cite{raiter2010} 
have constructed stellar evolution models 
that cover various stellar metallicities ($Z=0-1.0$ $Z_{\rm \odot}$), 
a variety of IMFs, and two star-formation histories of 
the instantaneous burst (burst SFH) and constant star-formation (constant SFH). 
These studies have theoretically examined the evolutions of 
spectral properties including emission lines for the stellar ages from 10$^{4}$ yr to 1 Gyr. 
From the theoretical computations, 
these studies have provided evolutions of  
$\beta$, EW$_{\rm 0}$(He{\sc ii}), and EW$_{\rm 0}$(Ly$\alpha$). 
To compute theoretical values of EW$_{\rm 0}$(Ly$\alpha$) and EW$_{\rm 0}$(He{\sc ii}), 
\cite{schaerer2003}  and \cite{raiter2010} have assumed Case B recombination. 
One of the advantages of the models of  \cite{schaerer2003}  and \cite{raiter2010}
is that the models have 
fine metallicity grids at an extremely low metallicity range. 
These fine metallicity grids are useful 
because large EW$_{\rm 0}$(Ly$\alpha$) LAEs are thought to have extremely low metallicities.  
Among the results of \cite{schaerer2003}  and \cite{raiter2010}, 
we use the predictions for six metallicities, 
$Z=0$ (Pop {\sc III}), $5\times10^{-6}$ $Z_{\odot}$, $5\times10^{-4}$ $Z_{\odot}$, 
$0.02$ $Z_{\odot}$, $0.2$ $Z_{\odot}$, and $1.0$ $Z_{\odot}$. 
We adopt three power-law IMFs, 
(A) the Salpeter IMF at the mass range of $1-100$ $M_{\odot}$,
(B) the top-heavy Salpeter IMF at the mass range of $1-500$ $M_{\odot}$,
and 
(C) the Scalo IMF (\citealt{scalo1986}) 
at the mass range of $1-100$ $M_{\odot}$. 
Table \ref{tab:models} summarizes the IMFs and their parameters. 

\begin{figure*}
\centering
\includegraphics[width=12cm]{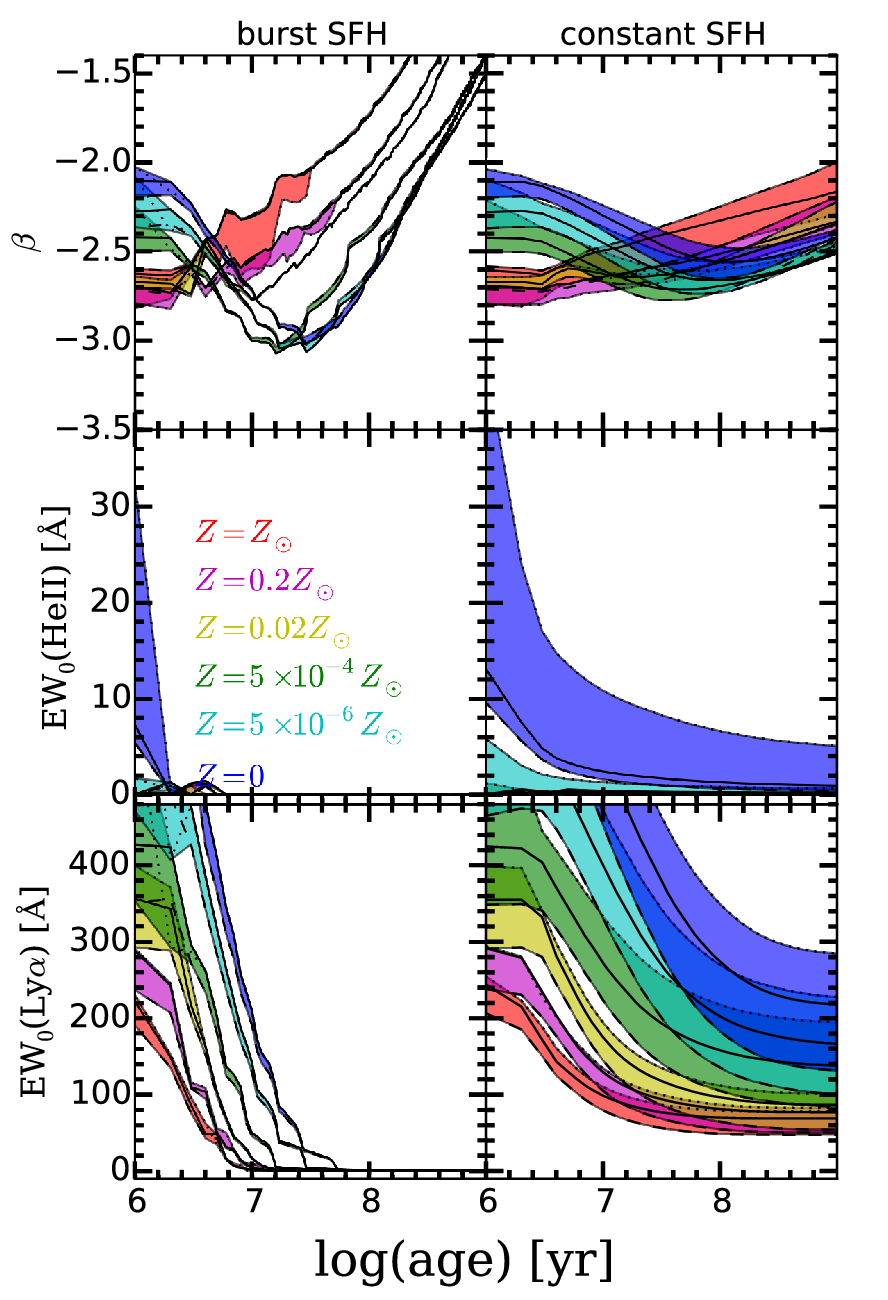}
\caption[]
{
Theoretical evolutions of $\beta$, EW$_{\rm 0}$(He{\sc ii}), and EW$_{\rm 0}$(Ly$\alpha$) values. 
The left and right panels are the results 
for the burst SFH and constant SFH, respectively. 
For each panel, 
the solid, dotted, and dashed lines correspond to the IMFs of A, B, and C, respectively
(Table \ref{tab:models}). 
The blue, cyan, green, yellow, magenta, and red regions denote evolution ranges 
traced by the IMFs for metallicities of 
$Z=0$, $5\times10^{-6}$ $Z_{\odot}$, $5\times10^{-4}$ $Z_{\odot}$, 
$0.02$ $Z_{\odot}$, $0.2$ $Z_{\odot}$, and $1.0$ $Z_{\odot}$, respectively. 
}
\label{fig:theoretical_ew_lya_heii_beta}
\end{figure*}

Figure \ref{fig:theoretical_ew_lya_heii_beta} plots the evolutions 
of $\beta$, EW$_{\rm 0}$(He{\sc ii}), and EW$_{\rm 0}$(Ly$\alpha$). 
The top panels of Figue \ref{fig:theoretical_ew_lya_heii_beta} are the $\beta$ evolutions. 
$\beta$ values are sensitive to the stellar and nebular continuum. 
The $\beta$ evolution for extremely low metallicity cases ($Z = 0-5\times10^{-4}$ $Z_{\odot}$) 
is significantly different from that for relatively high metallicity cases ($Z = 0.002-1.0$ $Z_{\odot}$). 
We explain the burst SFH case. 
In the relatively high metallicity cases ($Z = 0.002-1.0$ $Z_{\odot}$), 
the $\beta$ value monotonically increases as the stellar age increases. 
This is due to the fact that 
the dominant stellar continuum is red for old stellar ages. 
The value of $\beta \sim -2.7$ is expected at the very young stellar age 
of log(age yr$^{-1}$) $\sim 6.0-7.0$.  
In contrast, in the extremely low metallicity cases ($Z = 0-5\times10^{-4}$ $Z_{\odot}$), 
the $\beta$ value behaves as a two-value function. 
This is because the $\beta$ value 
is determined by both the stellar and nebular continuum 
at the extremely low metallicities cases. 
In these cases, 
the nebular continuum is very {\it red} for young stellar ages. 
Thus,at the very young stellar age of log(age yr$^{-1}$) $\sim 6.0-6.5$, 
the $\beta$ value is relatively large, $\beta \sim -2.3$, 
due to the balance between the red nebular continuum 
and the blue stellar continuum. 
The contribution of the red nebular continuum to the $\beta$ value 
becomes negligible at log(age yr$^{-1}$) $\gtrsim 7.0$ 
because of the rapid decrease of ionizing photons. 
Therefore, 
the $\beta$ value reaches down to $\beta \sim -3.0$ at log(age yr$^{-1}$) $\sim 7.0-7.5$, 
then monotonically increases. 
For the constant SFH, the evolution of $\beta$ is smaller than 
that of the burst SFH.

The second top panels of Figure \ref{fig:theoretical_ew_lya_heii_beta} show 
the EW$_{\rm 0}$(He{\sc ii}) evolutions. 
The EW$_{\rm 0}$(He{\sc ii}) value rapidly decreases as the metallicity increases: 
EW$_{\rm 0}$(He{\sc ii}) $> 5$ \AA\ is expected only for $Z < 5\times10^{-6}$ $Z_{\rm \odot}$. 
In the case of the burst SFH, the timescale for the He{\sc ii} line 
to be visible is short, log(age yr$^{-1}$) $\lesssim 7.0$. 
This timescale reflects the lifetime of extremely massive hot stars. 
Again, the evolution of EW$_{\rm 0}$(He{\sc ii}) is 
larger in the burst SFH than that of the constant SFH. 

The bottom panels of Figure \ref{fig:theoretical_ew_lya_heii_beta} indicate 
the evolution of EW$_{\rm 0}$(Ly$\alpha$). 
A high EW$_{\rm 0}$(Ly$\alpha$) value is expected for 
a young stellar age and a low metallicity. 
In the case of the burst SFH, the timescale for the Ly$\alpha$ line to be visible 
is log(age yr$^{-1}$) $\lesssim 7.5$. 
This reflects the lifetime of O-type stars. 
The maximum EW$_{\rm 0}$(Ly$\alpha$) value can reach 
EW$_{\rm 0}$(Ly$\alpha$) $\sim 800-1500$ \AA\ for the Pop {\sc III} metallicity.

\begin{table}
\centering
\caption{Summary of IMF Model Parameters}
\label{tab:models}
\begin{tabular}{cccccc}
\hline
{Model ID}  & {IMF} & {Line type} & {$M_{\rm low}$} & {$M_{\rm up}$} & {$\alpha$} \\
{}  &  {}  &  {}  & {($M_{\rm \odot}$)} & {($M_{\rm \odot}$)} &  {} \\
{(1)} & {(2)} & {(3)} & {(4)}  & {(5)}   & {(6)} \\
\hline\\ 
A & Salpeter & solid & $1$ & $100$ & $2.35$\\
\hline\\
B & Salpeter & dotted & $100$ & $500$ & $2.35$\\
\hline\\
C &  Scalo & dashed & $1$ & $100$ & $2.75$\\
\hline 
\end{tabular}
%
\begin{minipage}{100mm}
\begin{flushleft}
(1) Model ID;
(2) IMF; 
(3) Line style in Fig. \ref{fig:theoretical_ew_lya_heii_beta}; 
(4) and (5) \\ Lower and upper mass cut-off values; 
and 
(6) IMF slope value. 
\end{flushleft}
\end{minipage}
\end{table}

\subsubsection{Comparisons of the Observational Constraints with the Models} \label{subsubsec:age_metal_comparison}

\begin{figure*}
\centering
\includegraphics[width=15cm]{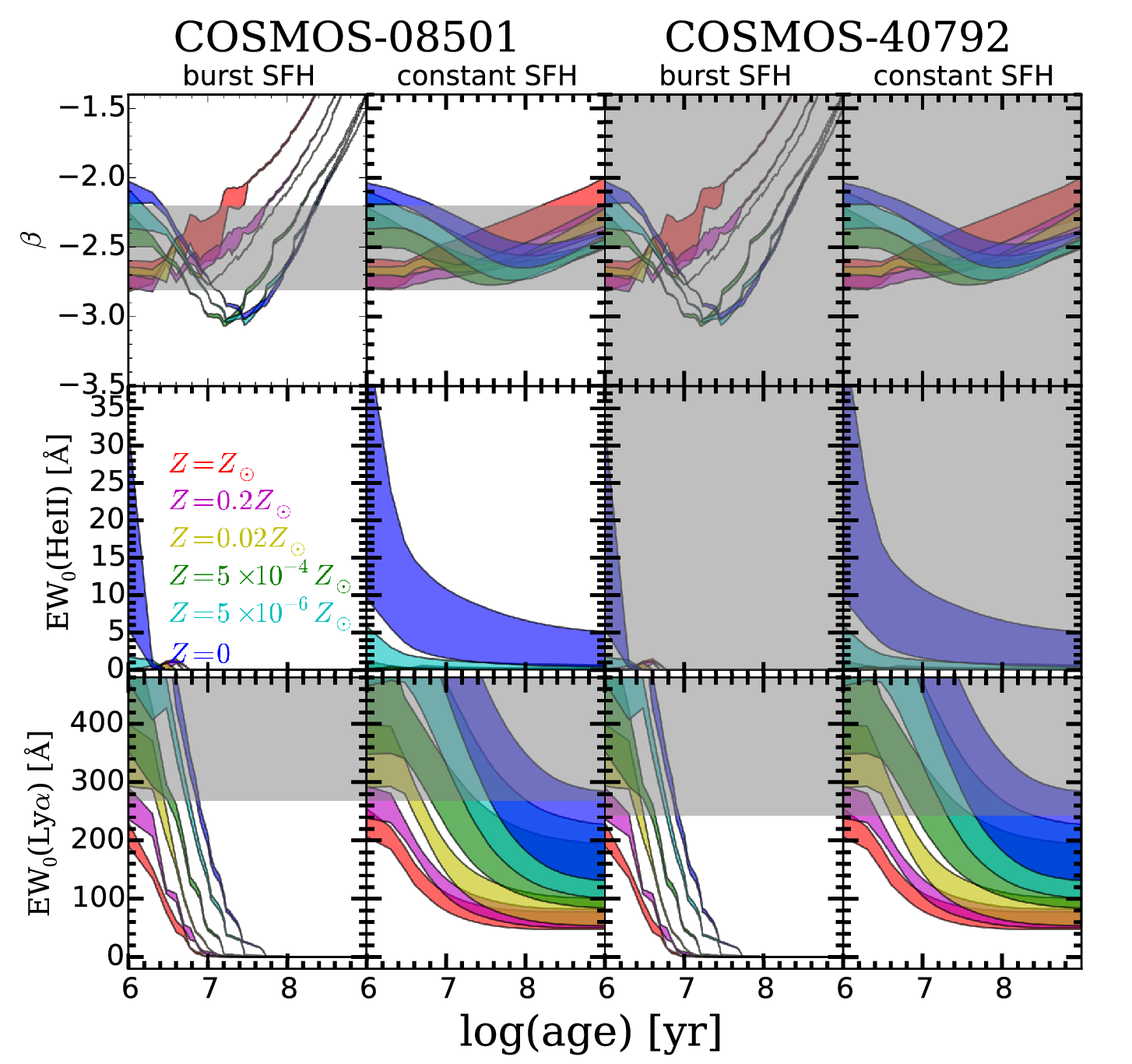}
\caption[]
{
Comparisons of the observational constraints of $\beta$, EW$_{\rm 0}$(He{\sc ii}), 
and EW$_{\rm 0}$(Ly$\alpha$) with the models 
for COSMOS-08501 and COSMOS-40792. 
For each object, the left and right panels show the results 
for the burst SFH and constant SFH, respectively.
The color codes for the different metallicities are the same as those in 
Fig. \ref{fig:theoretical_ew_lya_heii_beta}. 
The horizontal gray shaded regions indicate the ranges of the observed quantities. 
In the top panels, 
we plot intrinsic $\beta$ values, $\beta_{\rm 1800-2200}$ (\S \ref{subsec:estimate_ew_beta}), 
that are corrected for dust attenuation effects on $\beta$. 
In the bottom panels, we plot EW$_{\rm 0}$(Ly$\alpha$) values (Table \ref{tab:EW_beta_MUV})
as the lower limits of the intrinsic EW$_{\rm 0}$(Ly$\alpha$)  values. 
This is because the models of \cite{schaerer2003} and \cite{raiter2010} 
do not take into account the effects of Ly$\alpha$ scattering/absorption in the ISM and IGM. 
}
\label{fig:obs_ew_lya_heii_beta1}
\end{figure*}

\begin{figure*}
\centering
\includegraphics[width=15cm]{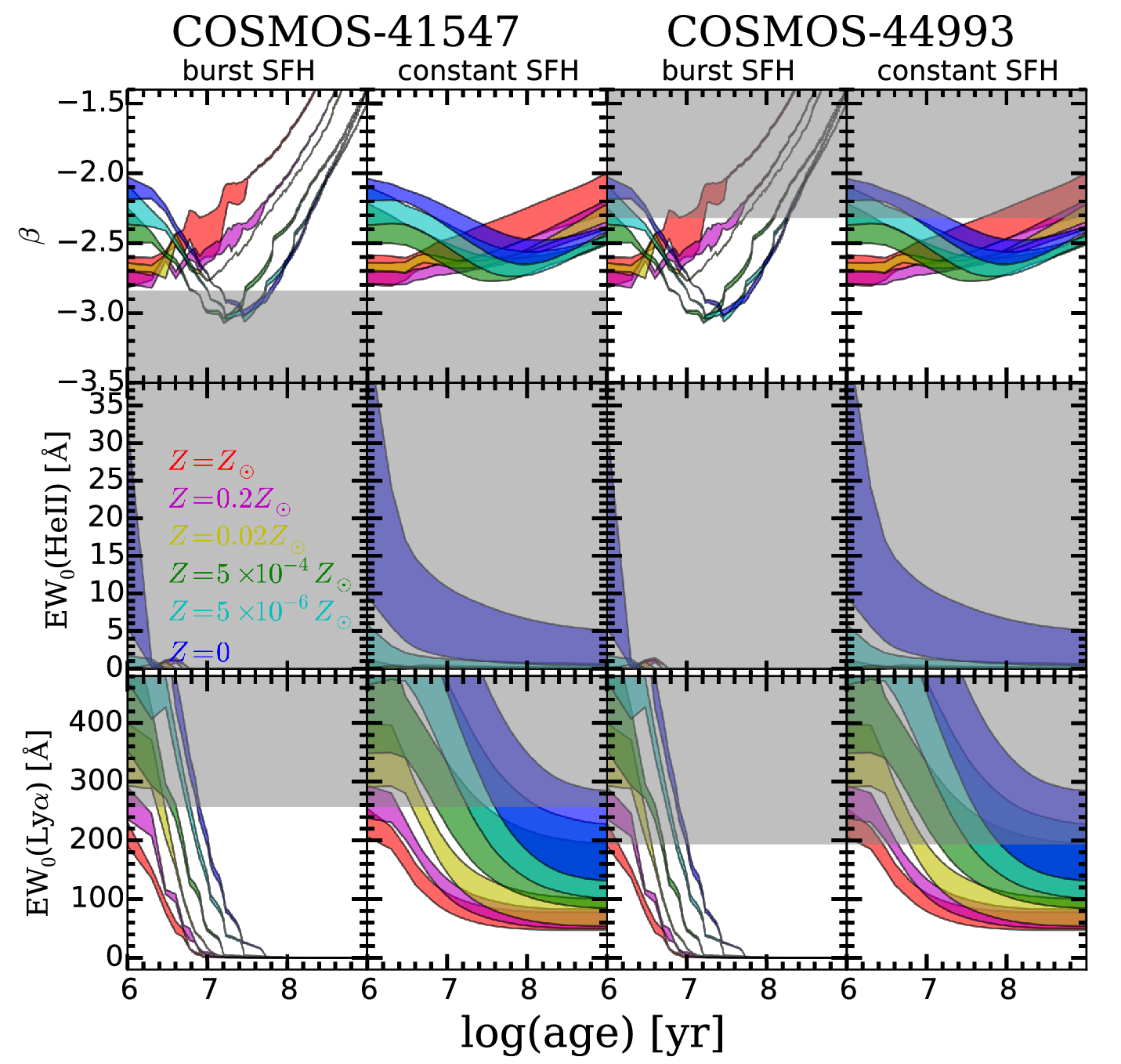}
\caption[]
{
Same as Fig. \ref{fig:obs_ew_lya_heii_beta1} 
for COSMOS-41547 and COSMOS-44993. 
}
\label{fig:obs_ew_lya_heii_beta2}
\end{figure*}

\begin{figure*}
\centering
\includegraphics[width=15cm]{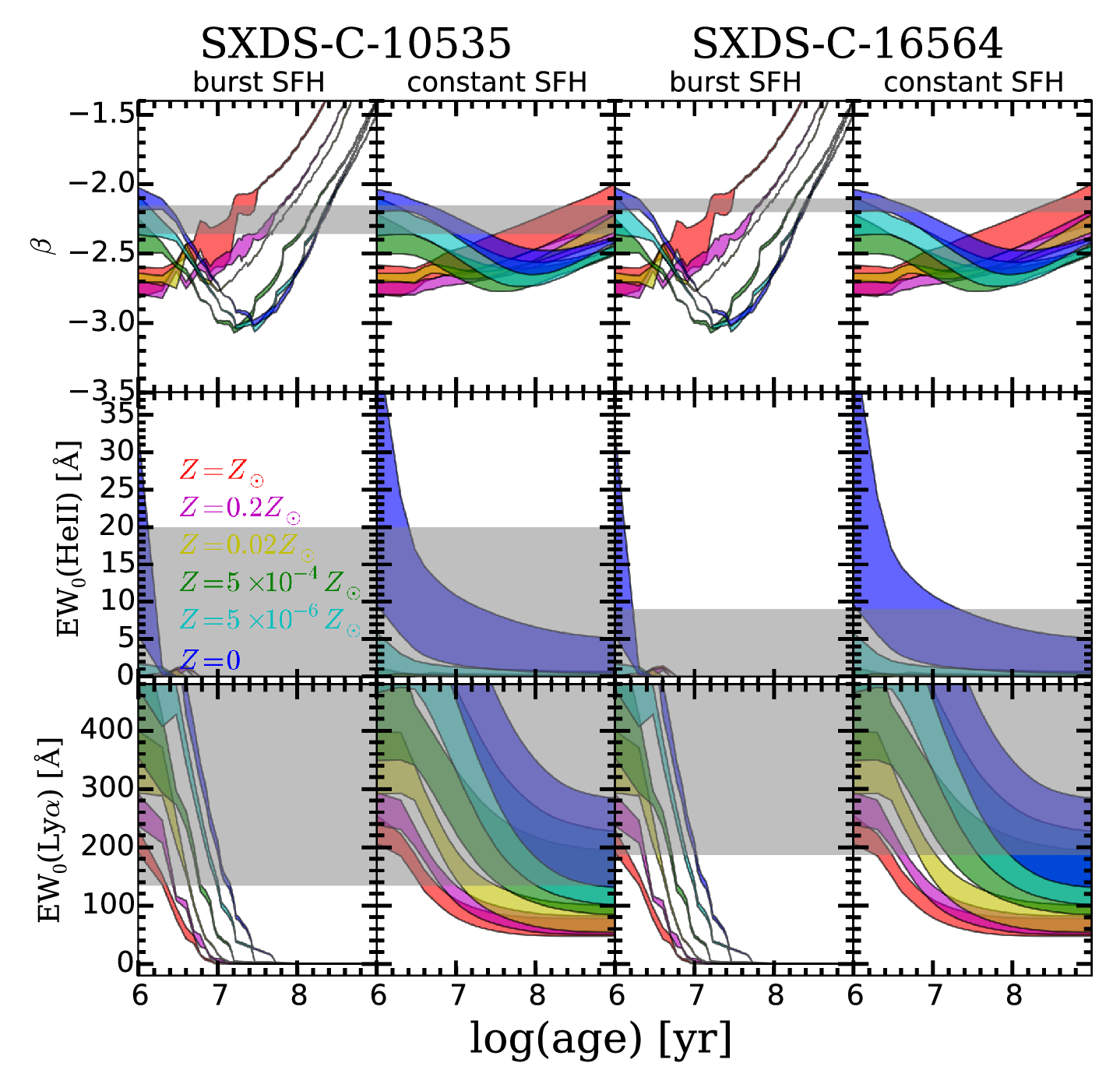}
\caption[]
{
Same as Fig. \ref{fig:obs_ew_lya_heii_beta1} for 
SXDS-C-10535 and SXDS-C-16564. 
}
\label{fig:obs_ew_lya_heii_beta3}
\end{figure*}

Figures \ref{fig:obs_ew_lya_heii_beta1} $-$ \ref{fig:obs_ew_lya_heii_beta3} 
compare the observational constraints of $\beta$, EW$_{\rm 0}$(He{\sc ii}), 
and EW$_{\rm 0}$(Ly$\alpha$) with the models. 
In these figures, gray shaded regions show the observed ranges 
of the three quantities.
In the top panels, we show the intrinsic UV continuum slopes, $\beta_{\rm1800-2200}$, 
for fair comparisons to the models (\S \ref{subsec:estimate_ew_beta}). 
In the second top panels, 
we present the upper limits of EW$_{\rm 0}$(He{\sc ii})
(\S \ref{subsec:upper_limit_helium}). 
The upper limits of EW$_{\rm 0}$(He{\sc ii}) are obtained except for COSMOS-08501. 
As can be seen, these values are not strong enough to 
place constraints on the stellar age and metallicity. 
In the bottom panels, it should be noted that 
the models of \cite{schaerer2003} and \cite{raiter2010} do not take into account 
the effects of Ly$\alpha$ scattering/absorption in the ISM and IGM. 
Thus, 
in Figures \ref{fig:obs_ew_lya_heii_beta1} $-$ \ref{fig:obs_ew_lya_heii_beta3}, 
we plot the EW$_{\rm 0}$(Ly$\alpha$) values (\S \ref{subsec:estimate_ew_beta}) 
as the lower limits of the intrinsic EW$_{\rm 0}$(Ly$\alpha$) values 
for fair comparisons to the models. 

\begin{figure*}
\centering
\includegraphics[width=13cm]{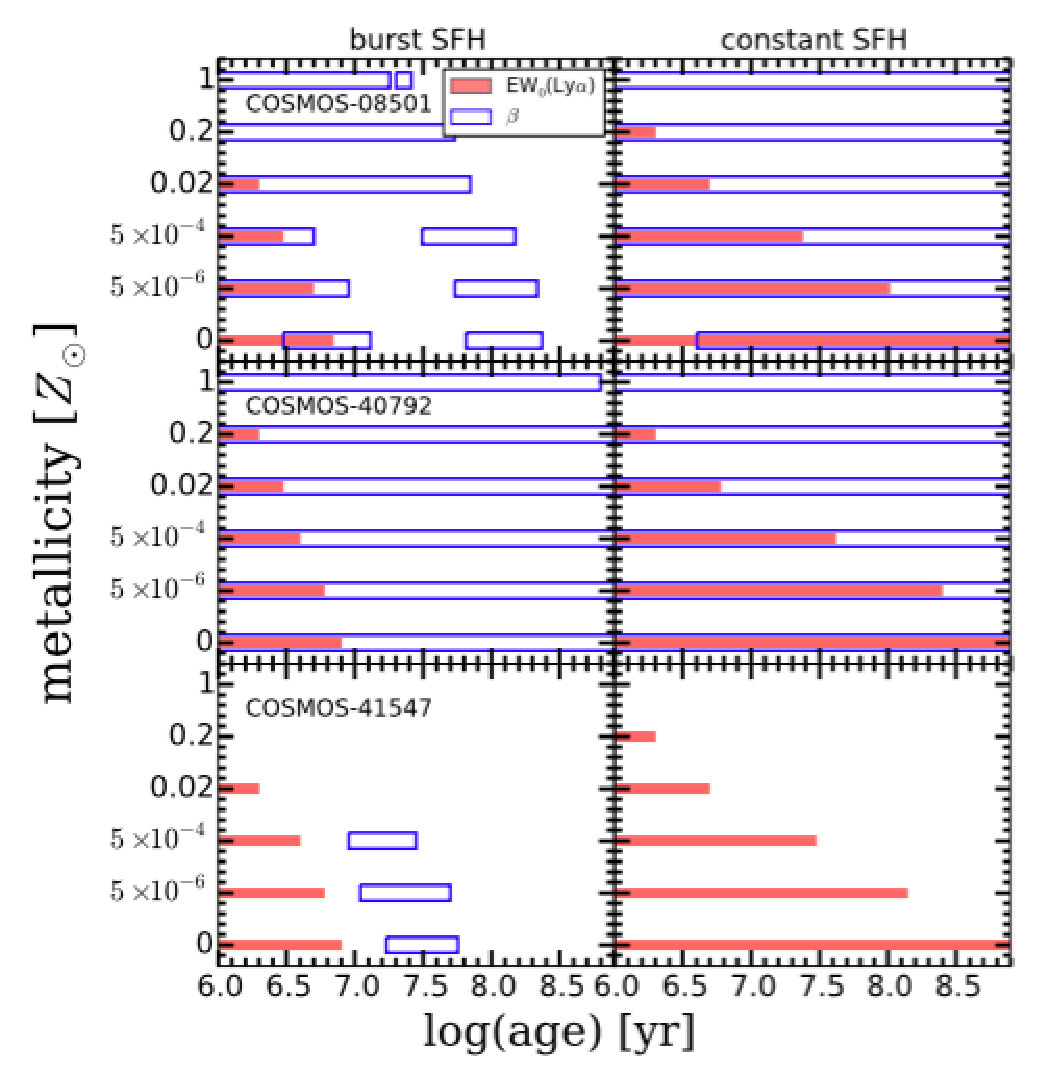}
\caption[]
{
Left and right panels show permitted ranges of the stellar age and metallicity 
in the burst SFH and constant SFH, respectively, 
for COSMOS-08501, COSMOS-40792, and COSMOS-41547. 
The red filled squares and blue open squares denote the permitted ranges 
derived from the EW$_{\rm 0}$(Ly$\alpha$) and $\beta$ values, respectively. 
The overlapped regions of the red filled squares and blue open squares 
are the final constraints on the the stellar mass and metallicity. 
}
\label{fig:age_metallicity1}
\end{figure*}

\begin{figure*}
\centering
\includegraphics[width=13cm]{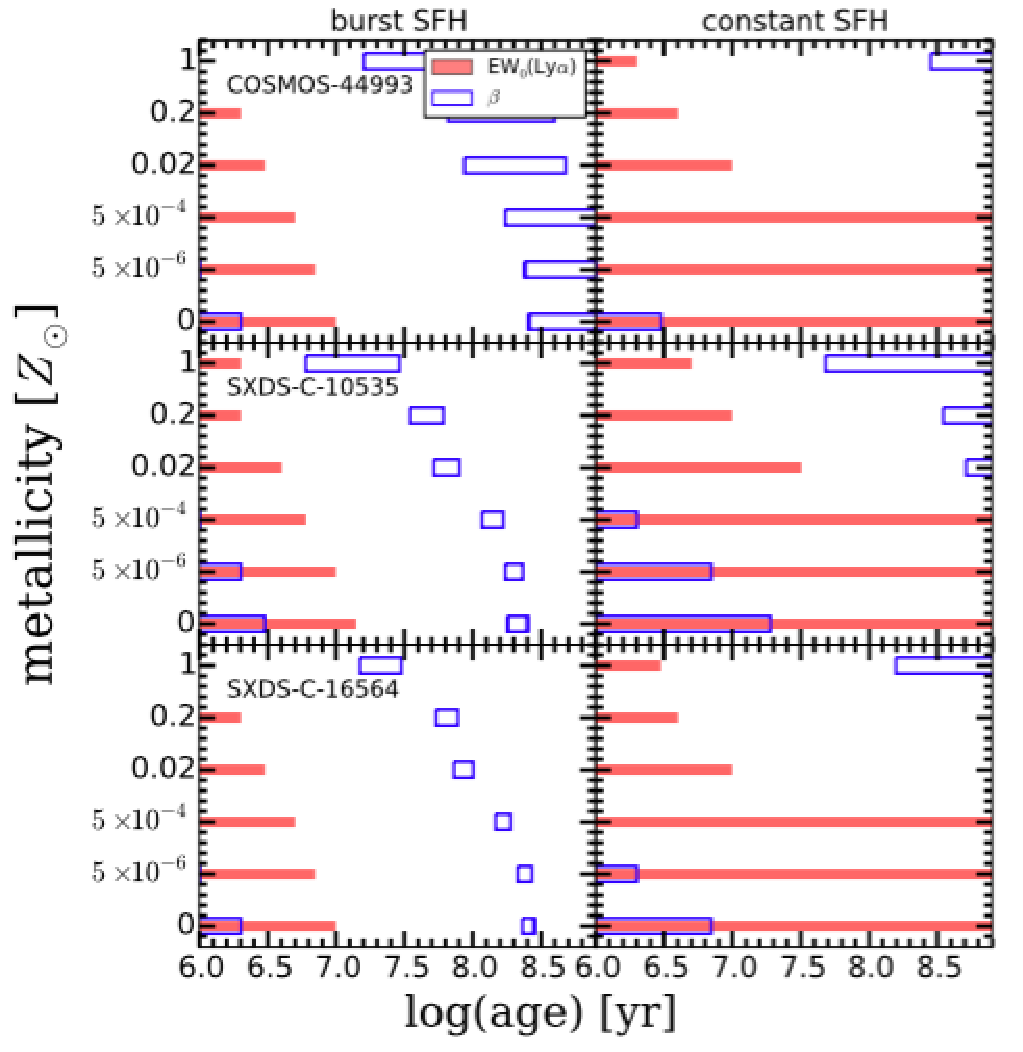}
\caption[]
{
Same as Fig. \ref{fig:age_metallicity1} for COSMOS-44993, SXDS-C-10535, 
and SXDS-C-16564. 
}
\label{fig:age_metallicity2}
\end{figure*}

In Figures \ref{fig:age_metallicity1} and \ref{fig:age_metallicity2}, 
we plot the two ranges of the stellar ages and metallicities 
given by the $\beta$ and EW$_{\rm 0}$(Ly$\alpha$) values. 
The overlapped ranges of the two are adopted as the stellar ages and metallicities. 
Figures \ref{fig:age_metallicity1} and \ref{fig:age_metallicity2} clearly demonstrate that 
the combination of $\beta$ and EW$_{\rm 0}$(Ly$\alpha$) 
is powerful to constrain the stellar age and metallicity. 
Table \ref{tab:sf_age_metal} summarizes 
the permitted ranges of the stellar ages and metallicities of our LAEs. 
In Table \ref{tab:sf_age_metal}, we find that 
our LAEs have generally low metallicities of $Z \lesssim 0.2 Z_{\rm \odot}$. 
Interestingly, it is implied that 
at least a half of our large EW$_{\rm 0}$(Ly$\alpha$) LAEs 
would have young stellar ages of $\lesssim 20$ Myr 
and very low metallicities of $Z < 0.02$ $Z_{\rm \odot}$ 
(possibly $Z \lesssim 5 \times10^{-4}$ $Z_{\rm \odot}$) 
regardless of the SFH. 
In Figure \ref{fig:age_metallicity1}, we cannot obtain the stellar age and metallicity 
that simultaneously satisfy the $\beta$ and EW$_{\rm 0}$(Ly$\alpha$) values 
of COSMOS-41547. 
This object has an exceptionally large correction factor 
for $\beta_{\rm obs.1800-2200}$, $-1.25^{+0.20}_{-0.35}$, 
compared to the median correction factor of $-0.1\pm0.2$ (\S \ref{subsec:estimate_ew_beta}). 
This is due to its large dust extinction value, 
$E(B-V)_{\rm *} = 0.25^{+0.04}_{-0.07}$ (Table \ref{tab:sed_fit}). 
Therefore, 
the stellar age and metallicity of COSMOS-41547 would be exceptionally affected 
by the systematic uncertainty discussed in \S \ref{subsec:estimate_ew_beta}.
In \S \ref{subsec:other_scenarios}, we consider other scenarios 
for the reason why the models have failed to constrain the stellar age
and metallicity of COSMOS-41547. 

\begin{figure}
\centering
\includegraphics[width=9cm]{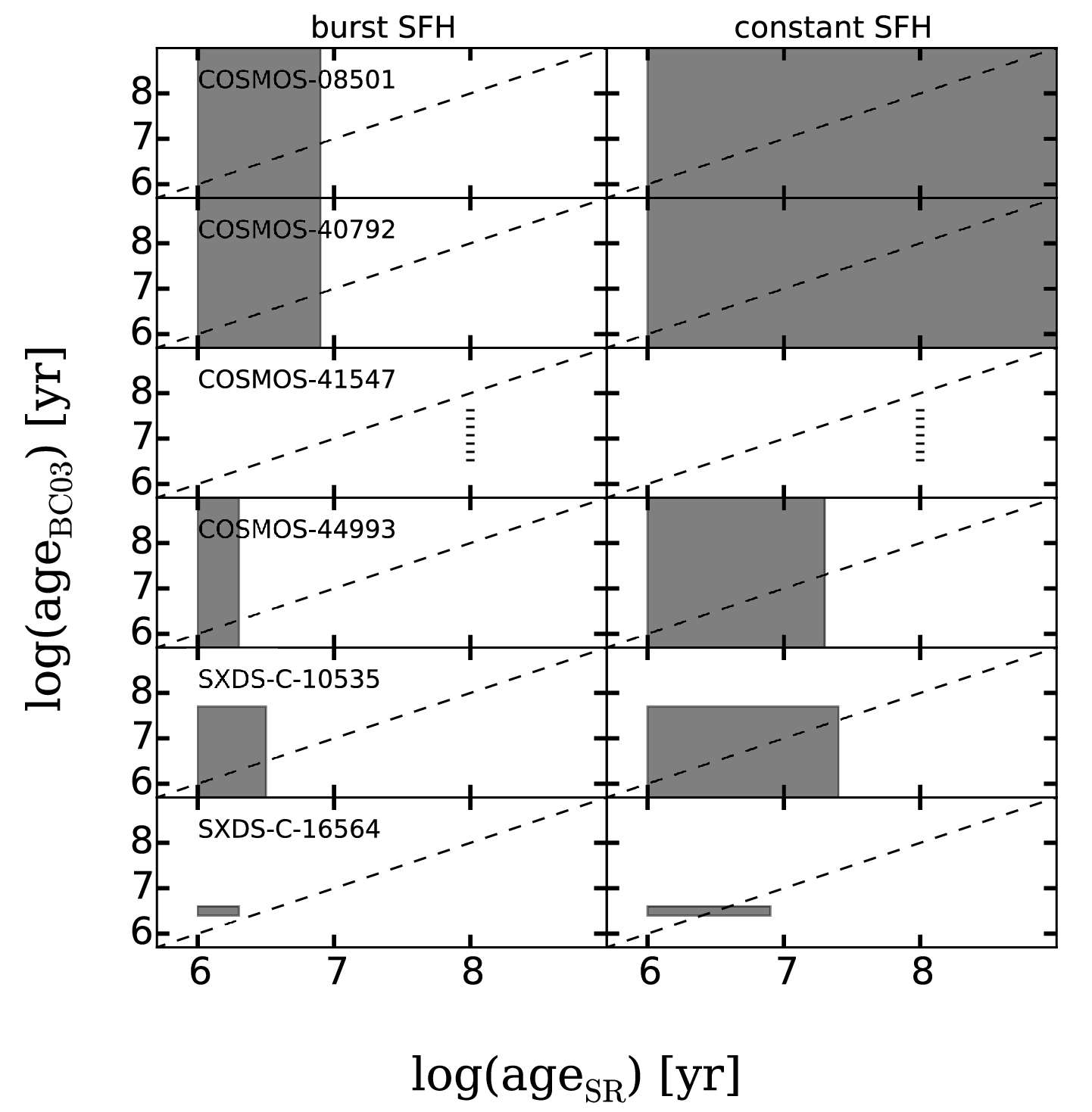}
\caption[]
{
Comparisons of the two stellar ages of our LAEs, 
the one derived from SED fitting, age$_{\rm BC03}$ (\S \ref{subsec:sed_fit}),
and the other from the comparisons of observables to the models of \cite{schaerer2003} and 
\cite{raiter2010}, age$_{\rm SR}$. 
For COSMOS-41547 whose age$_{\rm SR}$ cannot be constrained (see the text), 
the vertical dashed-line shows the range of age$_{\rm SED}$. 
In the left (right) panels, the age$_{\rm SR}$ values are obtained 
assuming the burst (constant) SFH. 
In each panel, the gray shaded region denotes the $1\sigma$ ranges of the two stellar ages. 
The dashed line indicates the one-to-one relation. 
}
\label{fig:age_age}
\end{figure}

Figure \ref{fig:age_age} compares the two stellar ages, 
the one derived from SED fitting (\S \ref{subsec:sed_fit}) 
and the other obtained with the models of \cite{schaerer2003} and \cite{raiter2010}.  
The former and the latter stellar ages are referred to as 
age$_{\rm BC03}$ and  age$_{\rm SR}$, respectively. 
The age$_{\rm BC03}$ value 
is determined by past star-formation activities. 
This is because the age$_{\rm BC03}$ value is estimated from the photometric data 
that cover the rest-frame optical wavelength. 
In contrast, 
the age$_{\rm SR}$ value represents the age of the most recent star burst activity. 
This is due to the fact that the age$_{\rm SR}$ value is obtained from the rest-frame UV data alone.
We find that the two stellar ages are consistent with each other 
within $1\sigma$ uncertainties regardless of the SFH. 
However, there is an exception, SXDS-C-16564 in the burst SFH case. 
In this case, the two stellar ages are 
consistent with each other within $2\sigma$ uncertainties. 
Among the two stellar ages, 
we adopt the age$_{\rm SR}$ values as the stellar ages of our LAEs. 
This is because the age$_{\rm SR}$ values 
are more realistic than the age$_{\rm BC03}$ values 
in the sense that the age$_{\rm SR}$ values are estimated 
with no assumption on the metallicity value. 

\begin{table*}
\centering
\caption{Summary of Stellar age and Metallicity of Our LAEs}
\label{tab:sf_age_metal}
\begin{tabular}{cccccccc}
\hline
{Object ID}  & {SFH} & {stellar age} & {} & {} & {} & {} & {}\\
{(1)}  & {(2)} & {(3)} & {} & {} & {} & {} & {}\\
\hline
{} & {} & {$Z/Z_{\rm \odot}$ = 0} & {$5\times10^{-6}$} & {$5\times10^{-4}$} & {0.02} & {0.2} & {1.0} \\
\hline 
COSMOS-08501 & burst & $3-8$ Myr & $<5$ Myr & $<3$ Myr & $<2$ Myr & $-$ & $-$\\
 & constant & $4-10^{3}$ Myr & $<100$ Myr & $<25$ Myr & $<5$ Myr & $<2$ Myr & $-$\\
\hline 
COSMOS-40792 & burst & $<8$ Myr & $<6.5$ Myr & $<5$ Myr & $< 3.5$ Myr & $<2$ Myr& $-$\\
 & constant & $<10^{3}$ Myr & $<250$ Myr & $<40$ Myr& $<6.5$ Myr & $< 2$ Myr& $-$\\
 \hline 
COSMOS-41547 & burst & $-$ & $-$ & $-$ & $-$ & $-$ & $-$ \\
 & const & $-$ & $-$ & $-$ & $-$ & $-$ & $-$ \\
 \hline 
 COSMOS-44993 & burst & $ <3.5$ Myr & $<2$ Myr & $-$& $-$ & $-$ & $-$ \\
 & constant & $<20$ Myr& $<8$ Myr& $<3$ Myr& $-$ & $-$ & $-$ \\
  \hline 
 SXDS-C-10535 & burst & $<3.5$ Myr & $<2$ Myr& $-$ & $-$ & $-$ & $-$ \\
 & constant & $<25$ Myr & $<10$ Myr& $<2$ Myr& $-$ & $-$ & $-$ \\
  \hline 
 SXDS-C-16564 & burst & $<2$ Myr & $-$ & $-$ & $-$ & $-$ & $-$ \\
 & constant & $<8$ Myr & $<2$ Myr & $-$ & $-$ & $-$ & $-$ \\
\hline
\end{tabular}
%
\begin{minipage}{170mm}
\begin{flushleft}
(1) Object ID;
(2) Star formation history; 
and 
(3) Permitted range of the stellar age for the each metallicity in the second row. 
\end{flushleft}
\end{minipage}
\end{table*}

\subsubsection{Limitations of Our Discussion} \label{subsubsec:age_metallicity_limitations}

We have derived the stellar ages and metallicities of our LAEs with two assumptions. 
First, we have presumed the Case B recombination. 
As pointed out by \cite{raiter2010} and \cite{dijkstra2014}, 
significant departures from Case B are expected 
at the low metallicity range of $Z \lesssim 0.03$ $Z_{\rm \odot}$ (see also \citealt{mas-ribas2016}).
The departures can contribute to strong Ly$\alpha$ emission 
up to EW$_{\rm 0}$(Ly$\alpha$) $\sim 4000$ \AA\ 
because of  (i) the increased importance of collisional excitation at the high gas temperature 
(ii) and the hard ionizing spectra emitted by metal poor stars (\citealt{dijkstra2014}). 
The departures can also contribute to weak He{\sc ii} emission 
compared to Case B (\citealt{raiter2010}). 
Thus, the constraints on the stellar age and metallicity may not be correct. 
Second, we have assumed a limited number of IMFs. 
\cite{raiter2010} have argued that large uncertainties remain in the shape of the IMF 
of the metal-poor or metal-free stars. 
Therefore, the constraints on the stellar age and metallicity 
suffer from uncertainties due to the shape of the IMF.

\subsection{Other Scenarios of the Large EW$_{\rm 0}$(Ly$\alpha$)} \label{subsec:other_scenarios}

We have studied properties of our LAEs 
assuming that all Ly$\alpha$ photons are produced by star forming activities. 
However, several other mechanisms can also generate Ly$\alpha$ photons. 
These include photoionization induced by 
(i) AGN activities (e.g.,  \citealt{malhotra2002, dawson2004}) 
or 
(ii) external UV background sources such as QSOs 
(QSO fluorescence: e.g., \citealt{cantalupo2005, cantalupo2012}). 
In addition, Ly$\alpha$ photons can be produced by collisional excitation 
due to 
(iii) strong outflows (shock heating: e.g., \citealt{taniguchi2000, mori2004,oti-floranes2012})
or 
(iv) inflows of gas into a galaxy  (gravitational cooling: e.g., \citealt{haiman2000, dijkstra2009, rosdahl2012}). 
These mechanisms can enhance the Ly$\alpha$ production, 
leading to large EW$_{\rm 0}$(Ly$\alpha$) values. 
Moreover, (v) if a galaxy has a clumpy ISM,  
where dust grains are shielded by H{\sc i} gas, 
EW$_{\rm 0}$(Ly$\alpha$) values can be apparently boosted. 
This is because Ly$\alpha$ photons are resonantly scattered 
on the surfaces of clouds without being absorbed by dust, 
while continuum photons are absorbed through dusty gas clouds 
(e.g., \citealt{neufeld1991, hansen_oh2006, kobayashi2010, laursen2013, gronke2014}). 
We examine these five hypotheses.

{\bf AGN activities:}
AGN activities can enhance EW$_{\rm 0}$(Ly$\alpha$). 
However, we have confirmed that our LAEs do not host an AGN 
both on the individual and stacked bases (\S \ref{subsec:agn}). 
The scenario is unlikely. 

{\bf QSO fluorescence:}
According to the result of \cite{cantalupo2005}, 
QSOs can photoionize the outer layer of the ISM of nearby galaxies, 
enhancing EW$_{\rm 0}$(Ly$\alpha$) of the nearby galaxy. 
We examine this hypothesis in two ways. 
First, we have confirmed that there are no QSOs around any of our LAEs. 
Second, as discussed in \cite{kashikawa2012}, 
objects with fluorescent Ly$\alpha$ often do not have stellar-continuum counterparts.
However, our LAEs clearly have stellar continuum counterparts (Table \ref{tab:photometry}). 
Therefore, we conclude that the QSO fluorescence hypothesis is unlikely. 

{\bf Shock heating:}
Shock heating caused by strong outflows can produce Ly$\alpha$ photons 
(\citealt{taniguchi2000, mori2004, oti-floranes2012}). 
In this case, the Ly$\alpha$ morphology is expected to be spatially extended (\citealt{haiman2000, taniguchi2015}). 
However, our LAEs have spatially compact Ly$\alpha$ morphologies (\S \ref{subsec:sample}). 
To obtain a definitive conclusion, 
it is useful to perform follow-up observations 
targeting [{\sc Sii}] and [{\sc Nii}] emission lines. 
This is because  [{\sc Sii}] and [{\sc Nii}] emission lines 
are sensitive to the presence of shock heating (e.g., \citealt{newman2012}). 
It is also interesting to perform follow-up observations 
targeting metal absorption lines. 
With blue-shifts of metal absorption lines with respect to the systemic redshifts, 
we can examine if outflow velocities are large enough 
to cause shock heating in our LAEs 
(e.g., \citealt{shapley2003, shibuya2014b, rivera-thorsen2015}).

{\bf Gravitational cooling:}
Ly$\alpha$ photons can be also generated by gravitational cooling. 
The gravitational binding energy of gas inflowing into a galaxy 
is converted into thermal energy, 
then released as Ly$\alpha$ emission. 
Ly$\alpha$ emission produced by gravitational cooling 
is predicted to be spatially extended (\citealt{rosdahl2012}). 
The compact Ly$\alpha$ morphologies of our LAEs 
do not favor the hypothesis. 
Deep H$\alpha$ data would help us to obtain a definitive conclusion. 
In the case of gravitational cooling, we expect 
a very high flux ratio of Ly$\alpha$ and H$\alpha$ lines, Ly$\alpha$/H$\alpha \sim100$ 
(\citealt{dijkstra2014}).
This high flux ratio can be distinguished from the ratio for the Case B recombination, 
Ly$\alpha$/H$\alpha = 8.7$. 

{\bf Clumpy ISM:}
Finally, the gas distribution of LAEs may not be smooth. 
\cite{duval2014} have theoretically investigated the condition of an ISM 
to boost EW$_{\rm 0}$(Ly$\alpha$) values. 
The EW$_{\rm 0}$(Ly$\alpha$) value can be boosted 
if a galaxy has an almost static (galactic outflows $<200$ km s$^{-1}$), clumpy, 
and very dusty ($E(B-V)_{\rm *} > 0.30$) ISM. 
The small median dust extinction value of our LAEs, $E(B-V)_{\rm *} = 0.02^{+0.04}_{-0.02}$, 
would be at odds with the hypothesis.

In summary, no clear evidence of the five scenarios has been found in our LAEs.

In \S \ref{subsubsec:age_metal_comparison}, 
we have shown that we cannot constrain the stellar age and metallicity of COSMOS-41547. 
One might think that the result is affected by e.g., a hidden AGN or collisional excitation. 
From Figure \ref{fig:obs_ew_lya_heii_beta2}, we have found  that we can 
constrain the stellar age and metallicity of this object 
if more than $60 \%$ of the observed Ly$\alpha$ flux is contributed 
from these additional mechanisms. 
If this is the case, we should see clear evidence of these effects. 
However, as we have shown, we do not see any clear evidence of these. 
Therefore, while the additional mechanisms could explain the failure of our method, 
the failure is most likely due to the systematic uncertainty as described 
in \S \ref{subsubsec:age_metal_comparison}.

\section{SUMMARY AND CONCLUSION} \label{sec:conclusions}

We have presented physical properties of spectroscopically confirmed LAEs 
with very large EW$_{\rm 0}$(Ly$\alpha$) values. 
We have identified six LAEs 
selected from $\sim3000$ LAEs at $z\sim2$ 
with reliable measurements of EW$_{\rm 0}$(Ly$\alpha$) $\simeq 200-400$ \AA\ 
given by careful continuum determinations 
with our deep photometric and spectroscopic data. 
These LAEs do not have signatures of AGN. 
Our main results are as follows. 

\begin{itemize}
\item 
We have performed SED fitting to derive physical quantities 
such as the stellar mass and dust extinction. 
Our LAEs have stellar masses of $M_{\rm *} = 10^{7-8} M_{\rm \odot}$
with a median value of $7.1^{+4.8}_{-2.8} \times 10^{7}M_{\rm \odot}$. 
The stellar masses of our LAEs is significantly smaller 
than those of small EW$_{\rm 0}$(Ly$\alpha$) LAEs at $z\sim2$, 
$M_{\rm *} = 10^{8-10} M_{\rm \odot}$ 
(\citealt{nakajima2012, oteo2015, shimakawa2016}). 
Our LAEs have stellar dust extinction values ranging 
from $E(B-V)_{\rm *} = 0.00$ to $0.25$ with a median value of 
$0.02^{+0.04}_{-0.02}$. 
The median value is lower than that of 
small EW$_{\rm 0}$(Ly$\alpha$) LAEs at $z\sim2$, $E(B-V)_{\rm *} = 0.2 - 0.3$ 
(\citealt{guaita2011, nakajima2012, oteo2015}).

\item 
By modeling FUV photometric data with no apriori assumption on $\beta$ values, 
we find that our LAEs have EW$_{\rm 0}$(Ly$\alpha$) values 
ranging from EW$_{\rm 0}$(Ly$\alpha$) $= 160$ to $357$ \AA, 
with a large mean value of  $252\pm30$ \AA. 
This confirms that LAEs with EW$_{\rm 0}$(Ly$\alpha$) $\gtrsim 200$ \AA\ exist. 
Our LAEs are characterized by 
the median values of $L({\rm Ly\alpha}) = 3.7\times10^{42}$ erg s$^{-1}$ 
and $M_{\rm UV} = -18.0$ 
as well as the small medan UV continuum slope of $\beta = -2.5\pm0.2$.

\item 
Using stellar masses and SFRs derived from SED fitting, 
we have investigated our LAEs' star-formation mode. 
With a high median sSFR ($\equiv$ SFR/$M_{\rm *}$) of $\sim100$ Gyr$^{-1}$, 
our LAEs typically lie above the lower-mass extrapolation 
of the SFMS $z\sim2$ 
defined by massive galaxies ($M_{\rm *} > 10^{10} M_{\rm \odot}$). 
An interpretation of the offset toward high sSFR 
is that our LAEs are in the burst star-formation mode. 
However, the offset can be also due to 
(i) a different slope of the SFMS at the low stellar mass range 
or (ii) a selection effect of choosing galaxies with bright emission lines (i.e., high SFRs)
at the low stellar mass range.

\item 
We have estimated the Ly$\alpha$ escape fraction, $f^{\rm Ly\alpha}_{\rm esc}$. 
For the three objects that have relatively small errors, 
the median value is calculated to be $f^{\rm Ly\alpha}_{\rm esc} = 0.68\pm0.30$. 
The high $f^{\rm Ly\alpha}_{\rm esc}$ value of our LAEs can be explained by  
the small dust content inferred from the small $E(B-V)_{\rm *}$ and $\beta$ values.

\item 
Our large EW$_{\rm 0}$(Ly$\alpha$) LAEs have 
a small mean FWHM$_{\rm int}$(Ly$\alpha$) of $212\pm32$ km s$^{-1}$, 
significantly smaller than those of small EW$_{\rm 0}$(Ly$\alpha$) LAEs 
and LBGs at the similar redshift. 
Combined with small EW$_{\rm 0}$(Ly$\alpha$) LAEs and LBGs in the literature, 
we have statistically shown that there is an anti-correlation 
between EW$_{\rm 0}$(Ly$\alpha$) and FWHM$_{\rm int}$(Ly$\alpha$). 
The small FWHM$_{\rm int}$(Ly$\alpha$) values of our LAEs 
can be explained either by  
(i) low $N_{\rm HI}$ values in the ISM, 
(ii)  low neutral-gas covering fractions of the ISM, 
or 
(iii) small dynamical masses.

\item 
We have placed constraints on the stellar ages and metallicities of our LAEs 
with the stellar evolution models of \cite{schaerer2003} and \cite{ raiter2010}. 
Our observational constraints of the large EW$_{\rm 0}$(Ly$\alpha$), the small $\beta$, 
and EW$_{\rm 0}$(He{\sc ii}) imply that at least half of our large EW$_{\rm 0}$(Ly$\alpha$) 
LAEs would have young stellar ages of $\lesssim 20$ Myr and 
very low metallicities of $Z < 0.02$ $Z_{\rm \odot}$ 
regardless of the SFH.

\item 
We have investigated five other scenarios of the large EW$_{\rm 0}$(Ly$\alpha$) values 
of our LAEs: 
AGN activities, QSO fluorescence, shock heating, gravitational cooling, 
and the presence of the clumpy ISM. 
Our sample does not show any clear evidence of these hypotheses. 

\end{itemize}

Among the results, 
the small $E(B-V)_{\rm *}$ and $\beta$ values are consistent with 
the high $f^{\rm Ly\alpha}_{\rm esc}$ values of our LAEs. 
The high $f^{\rm Ly\alpha}_{\rm esc}$ values are also 
consistent with the small FWHM(Ly$\alpha$) values indicative 
of the low {\sc Hi} column densities. 
We conclude that 
all of the low stellar masses, the young stellar ages, the low metallicities, 
and the high sSFR values are consistent with an idea 
that our large EW$_{\rm 0}$(Ly$\alpha$) LAEs represent 
the early stage of the galaxy formation and evolution 
with intense star-forming activities. 
The number of large EW$_{\rm 0}$(Ly$\alpha$) LAEs 
in this study is admittedly small. 
Hyper-Sprime Cam, a wide-field camera installed on Subaru, 
will be useful to increase the number of EW$_{\rm 0}$(Ly$\alpha$) LAEs 
at various redshifts. 

\section*{Acknowledgements}
We thank an anonymous referee for valuable comments 
that have greatly improved the paper. 
We are grateful to Alex Hagen and Giulia Rodighiero for kindly providing us 
with their data plotted in Figure \ref{fig:mass_sfr}. 
We thank Tohru Nagao, Ken Mawatari, Ryota Kawamata, and Haruka Kusakabe 
for their helpful comments and suggestions. 
In addition, we acknowledge the organizers of 
Lyman Alpha as an Astrophysical Tool Workshop  
at Nordita Stockholm in September 2013; 
this paper was enlightened by the talks and discussion that took place at that workshop.
This work was supported by 
World Premier International Research Center Initiative
(WPI Initiative), MEXT, Japan, 
and KAKENHI (23244022), (23244025), and (15H02064) 
Grant-in-Aid for Scientific Research (A)
through Japan Society for the Promotion of Science (JSPS).
T.H.  acknowledges the JSPS Research Fellowship for Young Scientists. 
K.N. was supported by the JSPS Postdoctoral Fellowships for Research Abroad.



\label{lastpage}

\end{document}